\documentclass[aps,pra,twocolumn, notitlepage,10pt,superscriptaddress,longbibliography]{revtex4-1}
\usepackage{graphicx}
\usepackage{times,bbm,amsmath,amssymb}
\usepackage{hyperref}
\usepackage{color}
\usepackage{xcolor}
\usepackage{graphicx}
\usepackage{dcolumn}%
\usepackage{bm}
\usepackage{soul}

\usepackage{nicefrac, xfrac}
\usepackage{amssymb}
\usepackage{pifont}

\usepackage{physics}

\newcommand{\parTitle}[1]{\noindent\emph{#1} --- }

\newcommand{\add}[1]{\textcolor{blue}{#1}}

\begin{document}

\title{Multiparameter quantum-enhanced adaptive metrology with squeezed light}

\author{Giorgio Minati}
\affiliation{Dipartimento di Fisica, Sapienza Universit\`{a} di Roma, Piazzale Aldo Moro 5, I-00185 Roma, Italy}

\author{Enrico Urbani}
\affiliation{Dipartimento di Fisica, Sapienza Universit\`{a} di Roma, Piazzale Aldo Moro 5, I-00185 Roma, Italy}

\author{Nicolò Spagnolo}
\affiliation{Dipartimento di Fisica, Sapienza Universit\`{a} di Roma, Piazzale Aldo Moro 5, I-00185 Roma, Italy}

\author{Valeria Cimini}
\email{valeria.cimini@uniroma1.it}
\affiliation{Dipartimento di Fisica, Sapienza Universit\`{a} di Roma, Piazzale Aldo Moro 5, I-00185 Roma, Italy}

\author{Fabio Sciarrino}
\affiliation{Dipartimento di Fisica, Sapienza Universit\`{a} di Roma, Piazzale Aldo Moro 5, I-00185 Roma, Italy}

\begin{abstract}

Squeezed light enables quantum-enhanced phase estimation, with crucial applications in both fundamental physics and emerging technologies. To fully exploit the advantage provided by this approach, estimation protocols must remain optimal across the entire parameter range and resilient to instabilities in the probe state.
In this context, strategies that rely on pre-calibrated squeezing levels are vulnerable to degradation over time and become sub-optimal when experimental conditions fluctuate. 
Here, we develop an adaptive multiparameter estimation strategy for ab-initio phase estimation, achieving sub-standard quantum limit precision in the full periodicity interval $[0,\pi)$, without relying on prior knowledge of the squeezing parameter. Our approach employs real-time feedback to jointly estimate both the optical phase and the squeezing level, ensuring robustness against experimental drifts and calibration errors. This self-calibrating scheme establishes a reliable quantum-enhanced sensing framework, opening new routes for practical scenarios and scalable distributed sensor networks using squeezed light.

\end{abstract}

\maketitle

\parTitle{Introduction}
Quantum-enhanced optical phase estimation using non-classical states of light \cite{pezze2014quantum} has attracted renewed interest in recent years, driven both by pioneering applications in the next-generation gravitational wave detectors \cite{acernese2019increasing, ligo2011gravitational, tse2019quantum} and by its impact on emerging areas such as imaging \cite{defienne2024advances}, spectroscopy \cite{adamou2025quantum, mukamel2020roadmap}, and distributed sensing \cite{zhuang2018distributed, gatto2019distributed, guo2020distributed}.
In this context, maximally entangled probes, such as $N00N$ states, provide, in principle, the highest phase sensitivity permitted by quantum mechanics \cite{polino2020photonic, barbieri2022optical}. However, their practical use is limited by the brightness of single-photon sources \cite{nagata2007beating} and by the detrimental effect of photon losses, which rapidly degrade their advantage \cite{demkowicz2012elusive}. To date, unconditional demonstrations of sub-standard quantum limit (SQL) performance with $N00N$ states have been restricted to the two-photon regime \cite{slussarenko2017unconditional}, making scalability to higher dimensions a major open challenge. As a result, the realization of quantum sensors based on such maximally entangled states remains difficult in practice.

Different routes have been explored, such as multi-pass strategies \cite{higgins2007entanglement, higgins2009demonstrating} and photon total angular momentum encoding \cite{cimini2023experimental}, but squeezed light provides a more scalable route to phase estimation beyond the SQL \cite{caves1981quantum, schnabel2017squeezed}, mitigating some of the key limitations associated with maximally entangled probes. Its effectiveness has already been demonstrated in the most recent upgrades of gravitational wave interferometers, which have shown the possibility of operating below the SQL for broadband signals, marking a milestone in the practical application of quantum resources for precision measurements \cite{jia2024ligo, acernese2023frequency, zhao2020frequency}.

\begin{figure*}[htb!]
\centering
\includegraphics[width=0.99\textwidth]{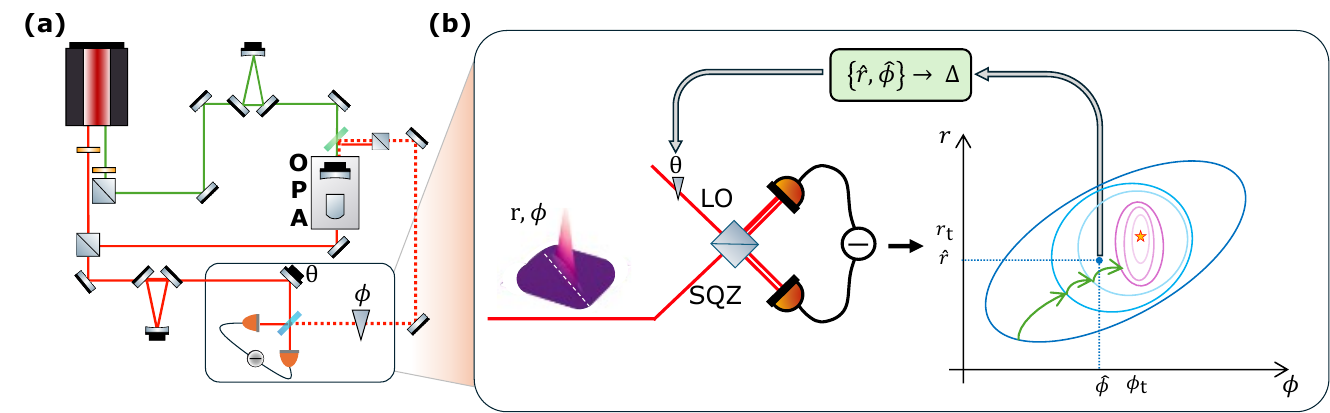}
\caption{\textbf{Sketch of the experimental setup and of the implemented protocol.} \textbf{(a)} Squeezed-vacuum probes are generated by pumping an optical parametric amplifier (OPA) with a single-mode continuous-wave laser at $532$ nm. To achieve high levels of squeezing, the pump and LO beams are both spatially filtered in identical mode-cleaner cavities to maximize the overlap between the squeezed field and the LO at the homodyne BS. The relative phase between the squeezed state and the LO is stabilized with a phase-locked loop (PLL). The LO phase $\theta$ is finely tuned by means of a piezoelectric stage, while an FPGA-based feedback system allows dynamic selection and locking of the measurement quadrature. The phase $\phi$ can be changed with a liquid crystal device in the path of the squeezed beam. \textbf{(b)} At each step of the protocol, the homodyne data collected are used to reconstruct the multiparameter posterior distribution. From this posterior, the estimate is updated and the optimal measurement angle $\theta$ is computed. A feedback $\Delta$, derived from the current estimates, is then applied to adjust the LO phase for the next measurement.
Repeating this cycle progressively refines the posterior until, at the end of the protocol, the estimates converge to the true values of phase $\phi_t$ and of the squeezing parameter $r_t$.}
\label{fig:setup}
\end{figure*}

Although squeezed light enables sensitivities below the SQL \cite{berni2015ab,nielsen2023deterministic, qin2023unconditional, pezze2008mach}, this is obtained at the expense of increased noise in the conjugate quadrature. Consequently, this enhanced precision is contingent upon a specific value of the optical phase, posing challenges for generic measurement strategies that necessitate the estimation of the phase as a fully unknown parameter from scratch. This limitation becomes critical in \emph{ab-initio} phase estimation, where no prior knowledge of the parameter is available. Therefore, in this regime, a measurement strategy based on fixed, pre-determined observables can not achieve optimal precision over the full parameter range without feedback. An approach to overcome this limitation is to implement adaptive strategies that permit to maintain consistent precision across the entire phase range accessible with the probe state, independently of the specific parameter values \cite{valeri2023experimental, berry2006adaptive,dinani2017adaptive,armen2002adaptive,wiseman1995adaptive, berry2002adaptive}.
Previous implementations of adaptive phase estimation with squeezed states have all relied on prior knowledge of the amount of squeezing in the probe \cite{berni2015ab, yonezawa2012quantum}, pre-calibrating the squeezing level, needed to optimize the measurement settings, limiting their applicability in practice. Moreover, they were also confined to half of the system periodicity.  

Realistic sensing scenarios introduce an additional layer of complexity. In many cases, not only is the sample-induced phase shift unknown, but properties of the probe state itself, such as the squeezing strength, may also be subject to variations during the measurement time \cite{roccia2018multiparameter, cimini2019adaptive, belliardo2024optimizing}. Addressing such situations requires a multiparameter estimation framework \cite{szczykulska2016multi, albarelli2020perspective}, in which both phase and probe parameters are inferred simultaneously. Since the corresponding observables are generally non-commuting, quantum incompatibility imposes trade-offs on precision \cite{albarelli2020perspective, pezze2025advances, albarelli2022probe, gessner2018sensitivity}, making it essential to investigate experimentally how adaptive protocols perform in this regime \cite{uola2016adaptive}. 
To this end, a stepwise estimation strategy has recently been introduced \cite{mukhopadhyay2025beating, sharma2025mitigating, mukhopadhyay2025saturable}, alongside several other methods aimed at boosting precision in this framework \cite{xia2023toward, hou2021zero, yuan2016sequential, liu2017control}.

In this work, we present a significant advancement in measuring unknown phase shifts with precision that surpasses the classical limit without requiring pre-calibration of the probe key parameters. By combining Homodyne Detection (HD) with Bayesian inference in an adaptive measurement procedure, we realize and experimentally validate an ab-initio multiparameter estimation scheme that simultaneously learns the unknown phase and relevant probe parameters. This calibration-free strategy demonstrates genuine quantum enhancement under realistic conditions, highlighting its technological relevance for robust, deployable phase sensing.
The implemented protocol dynamically adjusts the local oscillator (LO) phase in response to prior measurement outcomes, effectively steering the measurement basis toward the quadrature with the highest sensitivity at each step. Our approach achieves unconditional quantum-enhanced sensitivities for optical phase values across the entire periodicity range, without relying on systematic calibration of the probe. This is achieved through a multistep estimation procedure designed to overcome symmetrical constraints that previously confined investigations to the $[0,\pi/2]$ range. As a result, we extend phase estimation to the entire $[0,\pi)$ domain while directly inferring both the phase and the squeezing strength from the data. This article reports the experimental realization of quantum-enhanced optical phase estimation over the full $\pi$-range \cite{rodriguez2024adaptive}, marking a decisive step toward practical multiparameter estimation with squeezed light.

\parTitle{Theoretical framework} To benchmark the performance of our protocol, we refer to the parameter estimation framework \cite{d2000parameter}. In the single parameter scenario, the aim is to measure an unknown parameter $y$ by preparing a set of $M$ probe states, measure them according to a specific set of operators $\Pi_{k}$, and then process the data to retrieve information on the parameter. These choices determine the effective precision in the estimation process. For a chosen measurement strategy, the Fisher Information (FI) quantifies the information $F[y]$ about the unknown parameter $y$ contained in the measurement outcomes, optimized on all possible data processing strategies. This provides a first bound on the variance $\mathrm{Var}[y]$ for the unknown parameter, known as the Cramér–Rao bound (CRB) \cite{cramer1946mathematical, rao1945information}. This bound sets the achievable precision for a fixed probe state and measurement scheme. Note that the value of $F[y]$ can explicitly depend on the true value of the parameter. Optimizing the measurement strategies leads to the quantum Cramér-Rao bound (QCRB) \cite{helstrom1969quantum, gudder1985holevo, braunstein1994statistical}, which sets the ultimate precision associated with the specific probe state, and which is determined by the quantum Fisher Information (QFI) $F_{Q}$. Overall, the variance of any unbiased estimator satisfies the following chain of inequalities:
\begin{equation}
\label{eq:single_CRB}
    \mathrm{Var}[y] \overset{\mathrm{CRB}}{\geq} \frac{1}{M F[y]} \overset{\mathrm{QCRB}}{\geq} \frac{1}{M F_Q},  
\end{equation}
where $M$ represents the number of probe states used in the estimation. In our phase estimation scenario ($y \equiv \phi$), it has been shown \cite{monras2006optimal, olivares2009bayesian, berni2015ab} that for Gaussian probes, the optimal sensitivity to optical phase shifts corresponds to $F^{\mathrm{sq}}_Q = 2\sinh^2(2r)$, where $r$ is the squeezing parameter. Notably, for a sufficient level of squeezing, this surpasses the QFI of coherent states given by $F^{\mathrm{coh}}_Q = 4|\alpha|^2$  when considering an equivalent number of photons in the probe, since $|\alpha|^2$ and $\sinh^2(r)$ are the average photon number in the probe for each scheme. This limit can be saturated using HD combined with either maximum-likelihood or Bayesian estimation \cite{olivares2009bayesian}.

The parameter dependence of the FI highlights the need for adaptive measurement strategies, which dynamically adjust the detection basis to maintain optimal sensitivity across the entire phase range. In fact, fixed quadrature measurements provide sub-SQL precision only in restricted phase intervals. This naturally leads to use the FI as the fundamental figure of merit, where an adaptive feedback loop can be designed to steer the measurement toward the point of maximal FI, ensuring that the protocol operates close to the CRB across the entire accessible phase range.

In actual experimental implementations, it is also necessary to take into account that the effective probe state may also vary during the measurement process. Different phase shifts $\phi$ are associated with different fluctuations in the probe itself, such as variations in the amount of squeezing, thus affecting the attainable precision. As a result, the FI must be regarded as a function of both the unknown phase and the probe parameter $F[\vec{y}]$, where $\vec{y} = (\phi, r)$. Optimal strategies for phase estimation, therefore, can not be devised without simultaneously accounting for probe fluctuations, since relying on an incorrect pre-calibrated squeezing value can introduce systematic biases in the final estimate \cite{roccia2018multiparameter}. This naturally motivates the adoption of a multiparameter estimation framework, where the resources are invested in both phase and probe parameter estimation. In this multiparameter framework, the uncertainty in the simultaneous estimation of a set of parameters $\vec{y}$ is described by the covariance matrix $\bm{\Sigma}[\vec{y}]$, which is constrained by the multiparameter version of the (Q)CRB \cite{szczykulska2016multi}:
\begin{equation}
\label{eq:multi_CRB}
    \bm{\Sigma}[\Vec{y}] \overset{\mathrm{CRB}}{\succeq} \frac{1}{M} \bm{F}^{-1}[\vec{y}] \overset{\mathrm{QCRB}}{\succeq} \frac{1}{M} \bm{F_Q}^{-1}.  
\end{equation}
Here, $\bm{F}[\vec{y}]$ and $\bm{F_Q}$ denote the FI and QFI matrices, respectively, and the notation $\bm{A}\succeq\bm{B}$, indicates that $\bm{A}-\bm{B}$ is positive semidefinite. In this context, one must consider the fact that there is no guarantee that a single measurement strategy is optimal for estimating multiple parameters, meaning that, in general, is not ensured that the estimates can attain their QCRBs simultaneously \cite{ragy2016compatibility}. Furthermore, in the multiparameter scenario $\bm{F}[\vec{y}]$ can present singularities for some specific parameter values \cite{yang2025overcoming}. This is the case of joint estimation of the phase and squeezing parameter with a squeezed vacuum state, reflecting the fact that the measurement outcomes do not provide independent information about both parameters. From a practical perspective, these aspects make the design of adaptive protocols more demanding.

\parTitle{Experimental platform and adaptive protocol}
The squeezed vacuum state is generated at a wavelength of $1064$ nm through the cavity-enhanced interaction of a pump beam at $532$ nm with a periodically-poled nonlinear crystal inside an Optical Parametric Amplifier (OPA). The phase of the generated squeezed vacuum state is kept fixed in time using an FPGA-based locking scheme, and, via a liquid crystal device, we set the arbitrary phase $\phi$, which corresponds to the parameter of interest to be estimated. For this purpose, we perform homodyne measurements along an arbitrary quadrature angle, determined by the phase $\theta$ of the LO that can be controlled with a piezoelectric stage. The main components of the experimental setup are illustrated in Fig.\ref{fig:setup}. More details about the apparatus are reported in the SI.

\begin{figure}[ht!]
\centering
\includegraphics[width=0.99\columnwidth]{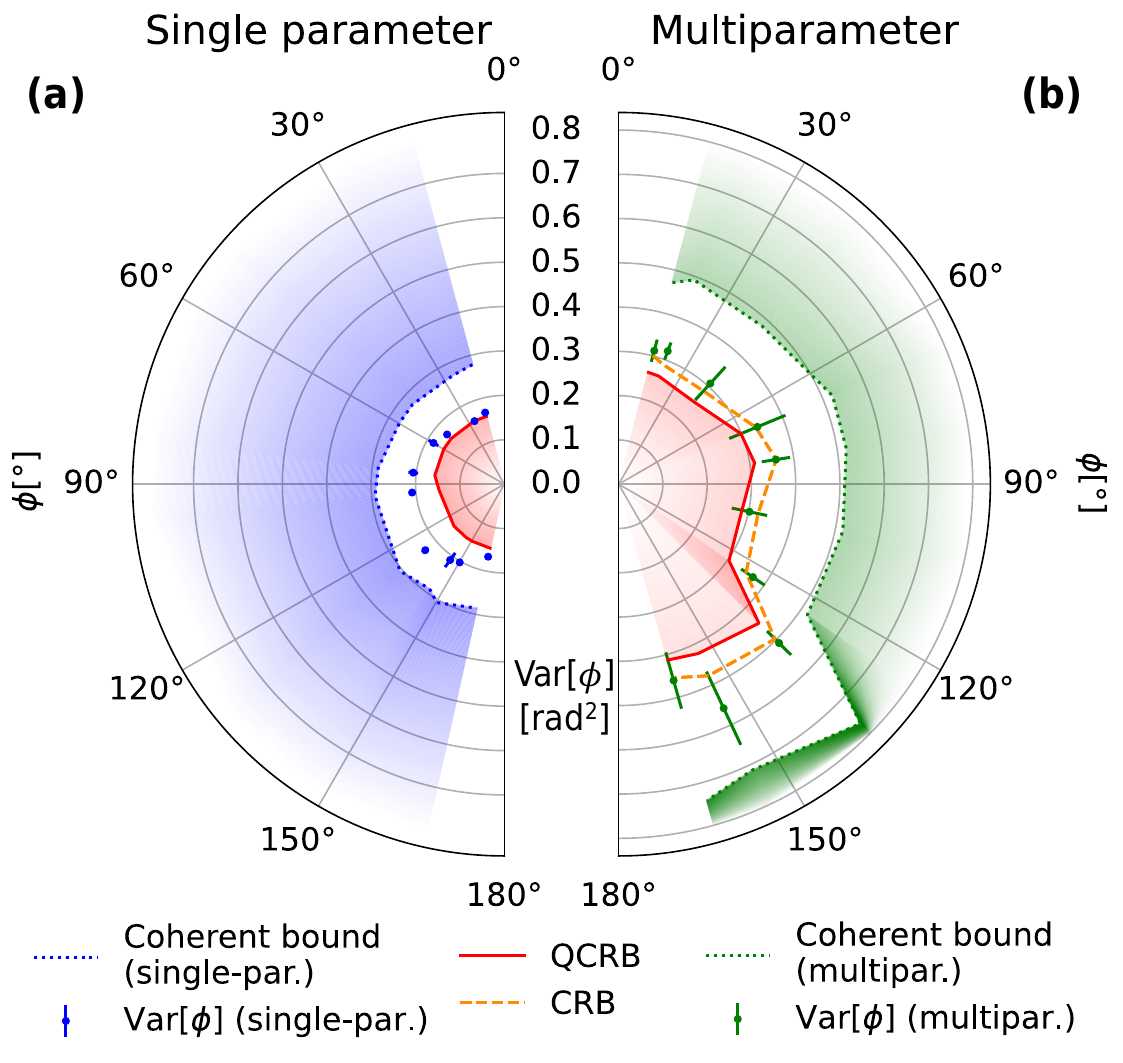}
\caption{\textbf{Single- and multiparameter variances of the phase estimation.} We report the variances of the adaptive phase estimation (radial axis) as a function of the phase $\phi$ (angular axis), spanning the entire range $[0, \pi)$. The experimental results $\mathrm{Var}[\phi]$ obtained in the single- [panel \textbf{(a)}] and multiparameter [panel \textbf{(b)}] adaptive protocols are reported as green and blue dots, respectively. The corresponding error bars represent the standard deviations over 5 (10) repetitions of the experiment, in the single-parameter (multiparameter) case. In both panels, we also report the corresponding classical bounds i.e., the QCRB for a coherent probe with the same average number of photons, as a blue line for the former and a green line for the latter case, while the QCRB is represented by a red line in both cases. For the multiparameter protocol we also report the actual CRB (orange dashed line).
The experimental single-parameter protocol employs a total of $M=5000$ homodyne measurements, among which $M_{\mathrm{R}}=200$ are used to obtain the rough estimate. Instead, in the multiparameter estimation, the total number of measurements amounts to $M=20000$, while the rough estimation needs $M_{\mathrm{R}} = 1200$ of them. 
}
\label{fig:results}
\end{figure}

For each unknown phase value $\phi$, we employ a Bayesian estimation strategy \cite{gebhart2021bayesian} to process the experimental data. Specifically, a sequence of $M$ homodyne measurements,
$X=\{x_1,x_2,...,x_M\}$, is collected and used to update our knowledge of the parameter through Bayes’ theorem \cite{olivares2009bayesian}.
Assuming no specific prior knowledge regarding both the parameter of interest $\phi$ and the squeezing level of the employed probe, i.e., adopting a flat prior distribution within the interval $\phi\in[0,\pi)$ and $r\in[0,3]$. The posterior probability reconstructed after a large number $M \gg 1$ of measurements can be expressed as:

\begin{equation}
p(\varphi,r|X) = \frac{1}{\mathcal{N}}\prod_{x\in X} p(x|\varphi,r)^{Mp(x|\varphi,r)}. 
\end{equation}
Here, $p(x|\varphi)$ represents the homodyne probability distribution, which consists of a Gaussian distribution centered in zero, and with variance:
\begin{equation}
    \sigma^2(\varphi,r,\eta) = \frac{1}{4}\big(\eta e^{-2r}\cos^2\varphi + (1-\eta) \cos^2\varphi + e^{2r}\sin^2\varphi\big),
\end{equation}
with $\varphi = \phi-\theta$ and $\eta$ accounting for losses in the setup. 

Once having reconstructed the posterior probability, we derive the estimates $\hat{\phi}$ and $\hat{r}$ of both parameters by calculating the mean value of the posterior distribution.  
To carry out ab-initio sub-SQL estimation of the phase $\phi$ over the full range $[0, \pi)$, we implement an online adaptive protocol consisting of two main stages. 
In the first step, we collect a small fraction of data $M_R$ for different settings of the LO phase $\theta$. In this way, we can remove the intrinsic ambiguity $\pi - \phi$ of squeezed-state interferometry, which constrained previous experimental phase estimation protocols with homodyne measurements \cite{berni2015ab} to $\phi \in [0, \pi/2]$. 
From these outcomes, a Bayesian update is performed to reconstruct the posterior distribution, from which rough estimates of both the phase and the squeezing parameter can be extracted. 
As the numerical evaluation of the posterior distribution can be computationally demanding, we approximate it by employing the Sequential Monte Carlo (SMC) technique \cite{granade2012robust, cimini2024benchmarking} that allows us to determine the online feedback. It consists of discretizing the posterior distribution into $n_{\mathrm{p}}$ particles $\{\phi_k\}_{k=1}^{n_p}$ and $\{r_k\}_{k=1}^{n_p}$, associated to weights $\{w_k(X, \phi)\}_{k=1}^{n_p}$ and $\{w_k(X, \phi,r)\}_{k=1}^{n_p}$, for the single and two-parameter protocols, respectively. The weights satisfy $\sum_k w_k = 1$ and are distributed according to the prior information. These are sequentially updated with the observed homodyne measurements $x\in X$, in such a way that the final mean and variance of the posterior can be efficiently computed with the following discrete sums:
\begin{eqnarray}
    \hat{\phi} &=&\sum_{k=1}^{n_\mathrm{p}} \omega_k(X,\phi-\theta)\phi_k, \\
    \hat{r} &=& \sum_{k=1}^{n_\mathrm{p}} \omega_k(X,r)r_k,\\
    \mathrm{Var}[\phi] &=& \sum_{k=1}^{n_\mathrm{p}} \omega_k(X,\phi-\theta)(\hat{\phi}-\phi_k)^2.
\end{eqnarray}
Further details on the SMC techniques are reported in the SI.
These preliminary estimates are then used to set the adaptive feedback by changing the LO phase at the most informative measurement point.

\parTitle{Results}
We start by benchmarking the protocol in the single-parameter estimation scenario, where the goal is to estimate only the phase $\phi$. In this framework, the first step of the adaptive procedure consists of acquiring data at two fixed values of the LO phase, i.e., $\theta = \{0, \pi/4\}$. This first batch of measurements provides a first rough estimate of the parameter, which in turn allows us to identify the optimal measurement projection and thus setting the phase of the adaptive protocol where the FI is maximum, thereby optimizing the next measurement projections.
For a pre-calibrated squeezing level $r$ and a transmission coefficient $\eta$, we can retrieve the effective squeezing parameter that captures the combined impact of squeezing and loss given by:

\begin{equation}
r_{\text{eff}}=\frac{1}{2}\log\big[\frac{\sigma^2_{\mathrm{asqz}}(r,\eta)}{\sqrt{\sigma^2_{\mathrm{asqz}}(r,\eta)\sigma^2_{\mathrm{sqz}}(r,\eta)}}\big].    
\end{equation}
The optimal measurement configuration corresponds to $ \phi_{\mathrm{opt}} = \frac{1}{2}\arccos{(\tanh{(2r_{\text{eff}})})}$. 
In practice, the measurement basis is aligned by choosing $\theta = \hat{\phi} - \phi_{\mathrm{opt}}$, where $\hat{\phi}$ is the current Bayesian estimate. After this step, a new block of quadratures is collected and used to update the posterior distribution, thereby updating the parameter estimates and yielding a more accurate determination of the optimal LO phase. This adaptive cycle is repeated three times, using in total the remaining $M-M_R$ measurements. At each iteration, the increasing amount of data yields sharper posteriors and progressively improves the determination of the adaptive measurement setting.

The experimental precision is obtained as the variance of the reconstructed posterior distribution for different values of $\phi$, after the collection of $M$ homodyne quadratures. The obtained results for the phase estimation in the entire range $\phi \in [0, \pi)$ are reported in Fig.\ref{fig:results}a. The experimental measurements are compared with the QCRB computed for squeezed-vacuum probes: 
$\mathrm{Var}_{\mathrm{sqz}}[\phi] = 1/F_{Q}^{\mathrm{sq,eff}}$, in red, representing the ultimate sensitivity attainable with Gaussian resources for the given squeezing level and loss in such conditions, with $F_{Q}^{\mathrm{sq,eff}} = 2 \sinh^2(2 r_{\mathrm{eff}})$.
The achieved variances are compared with the optimal precision attainable with classical resources of equal mean photon number, quantified by the corresponding QCRB for coherent states: $\mathrm{Var}_{\mathrm{coh}}[\phi] = 1 / (4 \langle{n}\rangle) = 1/[4 \sinh^2{(r)}]$ reported as a blue line in the figure. Importantly, to assess the unconditional advantage over classical light, the coherent bound is not loss-corrected but is evaluated at the probe actual mean photon number linked to the squeezing parameter $r$.
The bound is evaluated across all inspected phase values, each time accounting for the corresponding pre-calibrated levels of squeezing. These pre-calibrations reveal that the parameters are not fixed but instead fluctuate due to experimental instabilities, such as variations in the source during different data acquisitions and imperfections in the alignment procedures.

The experimental variances lie consistently below the classical bound, demonstrating unambiguous sub-SQL sensitivity. Moreover, the measured precision closely approaches the squeezed-state QCRB, confirming that our adaptive Bayesian strategy operates near the optimal point across the entire phase domain. 
While this strategy enables the saturation of the ultimate precision bound for the phase estimation with homodyne measurements, it also requires a precise calibration of the setup efficiency $\eta$ and of the squeezing level $r$, before performing the actual experiment. For this calibration to be effective, it must be sufficiently precise, hence, employing a significant amount of resources, which are not accounted in the overall resource budget. 

\begin{figure*}[htb!]
\centering
\includegraphics[width=0.99\textwidth]{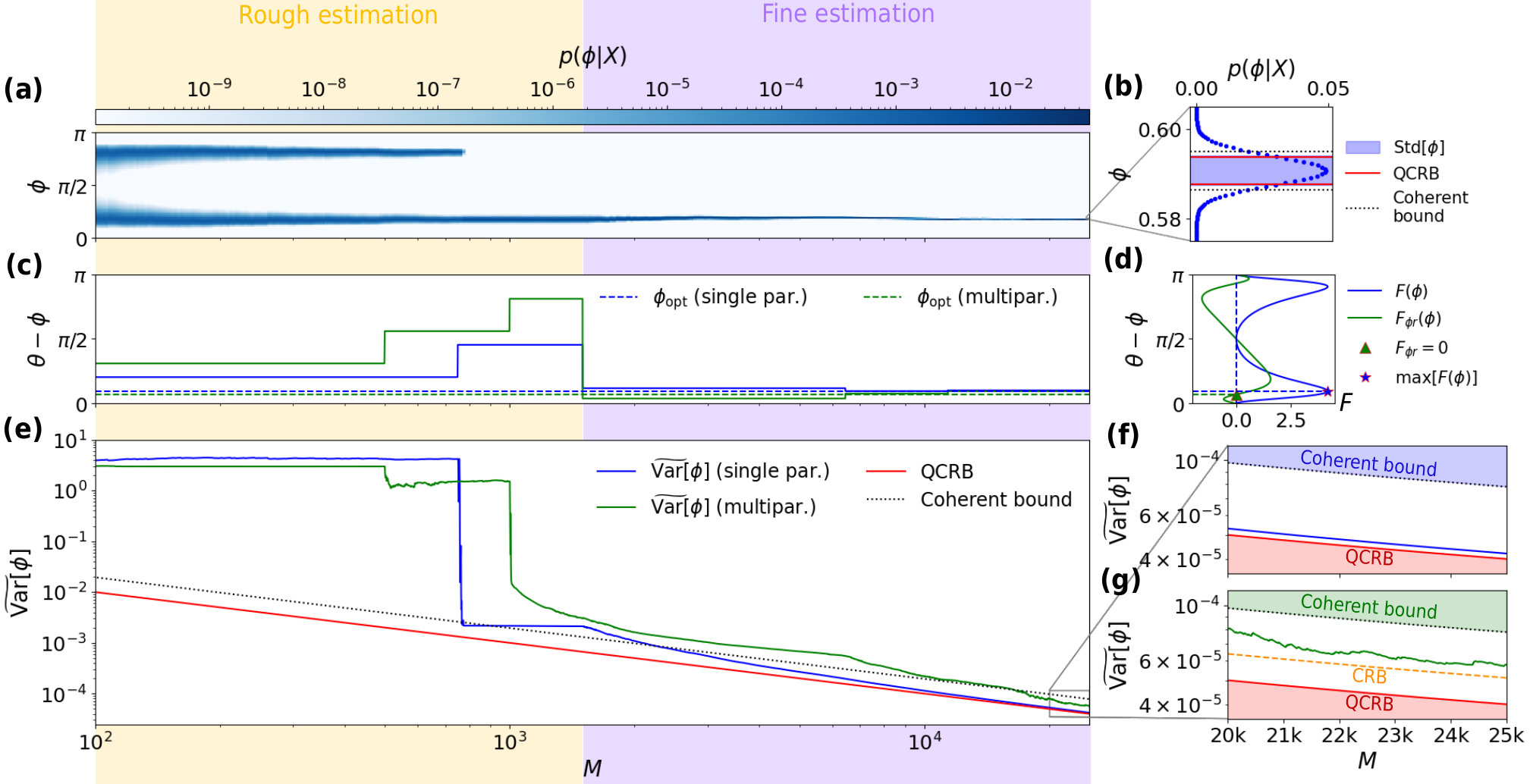}
\caption{\textbf{Adaptive phase estimation single- and multi-parameter protocols.} In panel \textbf{(a)}, we report the evolution of the posterior distribution $p(\phi|X)$ as a function of the number of homodyne data measured along the adaptive single-parameter protocol in the range $M\in[100,25000]$. During the rough estimation, highlighted by a yellow background, we observe that measuring different quadrature angles ($\theta=\{0,\pi/4\}$ and $\theta=\{0,\pi/4,\pi/2\}$ in the single- and multiparameter estimations, respectively) allows us to disambiguate the estimate between the ranges $\phi \in [0,\pi/2]$ and $\phi \in [\pi/2, \pi)$. At the end of this stage, the current estimation $\hat{\phi}$ is employed to shift the LO phase to its optimal value $\theta = \hat{\phi} - \phi_{\mathrm{opt}}$. 
The remaining measurements (purple background) are taken within the optimal homodyne configuration, which is further updated two times before reaching the final estimation, extrapolated from the posterior distribution reported in panel \textbf{(b)}. This inset shows how the precision of the phase estimation (blue shaded area), tightly approaches the squeezed QCRB (red line), while surpassing the classical bound (black dotted line). The changes of the LO angle, along the adaptive protocol, are illustrated in panel \textbf{(c)}. It shows that the quantity $\theta - \phi$ is close to its optimal value, depending on the chosen protocol. In the single parameter estimation (in blue), it concides with the value corresponding to the maximum of the FI [in panel \textbf{(d)}], while in the multiparameter approach (green), we aim at canceling the off-diagonal element of the FI matrix $F_{\phi r}$ [shown in panel \textbf{(d)}]. 
In panel \textbf{(e)}, the normalized variance $\widetilde{\mathrm{Var}}[\phi] \equiv \mathrm{Var}[\phi]\cdot F_{Q}^{\mathrm{sq,eff}}$ is reported as a function of $M$, thus having a common ultimate precision bound (solid red line) for both the single- and multiparameter variances, illustrated as blue and green lines, respectively. This panel also shows how the disambiguation and the LO feedback positively affect the precision of both the estimation protocols, although the multiparameter one exhibits a slower convergence to values below the classical bound (black dotted line).  Panels \textbf{(f)} and \textbf{(g)} provide zoomed-in views of the marked area, demonstrating convergence to the bound for the single-parameter strategy and for the multi-parameter strategy, respectively. The latter converges to the CRB computed for the two-parameter estimate that does not coincide with the QCRB. 
}
\label{fig:adaptivity}
\end{figure*}

Importantly, our experimental results show that in realistic scenarios, both the level of squeezing and losses are not fixed, but change depending on the specific phase value under investigation. This can arise because the liquid crystal may exhibit different transmission at different phases, or because the squeezing parameter drifts over time due to temporal instabilities of the source. 
Consequently, different phase values refer to different values of the bound on the attainable precision as emerges in Fig.\ref{fig:results}\add{a}. This observation further motivates the need of a multiparameter strategy, in which both the phase and the squeezing parameter are inferred simultaneously from the same dataset. Such an approach enables the protocol to adapt in real-time to fluctuations of the probe, eliminating the dependence on previous calibration, which can jeopardize the final estimate, while still achieving sub-SQL estimation of the phase, as described below.
We stress that within each estimation run, with the employed estimation model, we do not aim to track parameter variations dynamically; indeed, on the timescale of a few seconds, the phase-locking loop and the source stability are sufficient to maintain approximately constant values of $\phi$ and $r$.

In the multiparameter regime, for a fixed homodyne setting, the classical FI matrix becomes singular, which in this case showcases the impossibility of estimating simultaneously both parameters in the absence of prior information. The adaptive strategy, by measuring in different homodyne settings, restores a non-singular FI matrix, thus circumventing such a limitation. However, the measurement settings that maximize the phase sensitivity coincide with the sensitivity minimum for the squeezing parameter, and vice versa. This trade-off reflects the incompatibility of the relevant observables and precludes simultaneous attainment of the ultimate precision bounds for both parameters (see SI for additional details).

Since our primary goal remains to estimate $\phi$ with the highest possible precision, thus prioritizing the estimate of $\phi$ over $r$, it might be expected that selecting the single-parameter phase-optimal setting would suffice. In practice, however, an excessively imprecise estimate of $r$ degrades the adaptive LO feedback and therefore the phase estimate (see simulation results in SI).
To balance these competing requirements, we implement an LO feedback rule that aims to cancel the correlation between the two parameters' estimation errors.
The optimal strategy corresponds to set the quadrature where phase and squeezing estimates are effectively decorrelated, i.e., where the off-diagonal elements of the FI matrix $F_{\phi r}$  vanish. This corresponds to setting $\theta = \hat{\phi} - \arccos{\frac{e^{2\hat{r}}}{\sqrt{e^{4\hat{r}}+\eta}}}$, with $\eta$ the detection efficiency. Both the estimates $\hat{r}$ and $\hat{\phi}$, needed to compute the feedback updates, are retrieved as mean values of the two-dimensional posterior distribution, which is efficiently reconstructed with the SMC approach, further justifying its adoption to significantly speed up the numerical reconstruction, especially in this multiparameter framework. The precisions achieved on the phase estimates for the multiparameter protocol are reported in  Fig.\ref{fig:results}b, showing also in this case the ability to obtain quantum-enhanced performance, while ensuring robustness to a varying squeezing level. 
Overall, we observe gains of $1.79 \pm 0.40$ dB below the SQL, not corrected for loss and averaged over 10 different phase values, with the conventional single-parameter (pre-calibrated) approach and $1.78 \pm 0.33$ dB below the SQL with our ab-initio multiparameter strategy concerning the phase estimate. 
These results highlight both the robustness and the unconditional character of the protocol, since the quantum advantage is maintained without prior calibration of the probe and for all the values of $\phi$, overcoming also the restrictions of previous approaches limited to narrower phase intervals.

\begin{figure}[htb!]
\centering
\includegraphics[width=0.95\columnwidth]{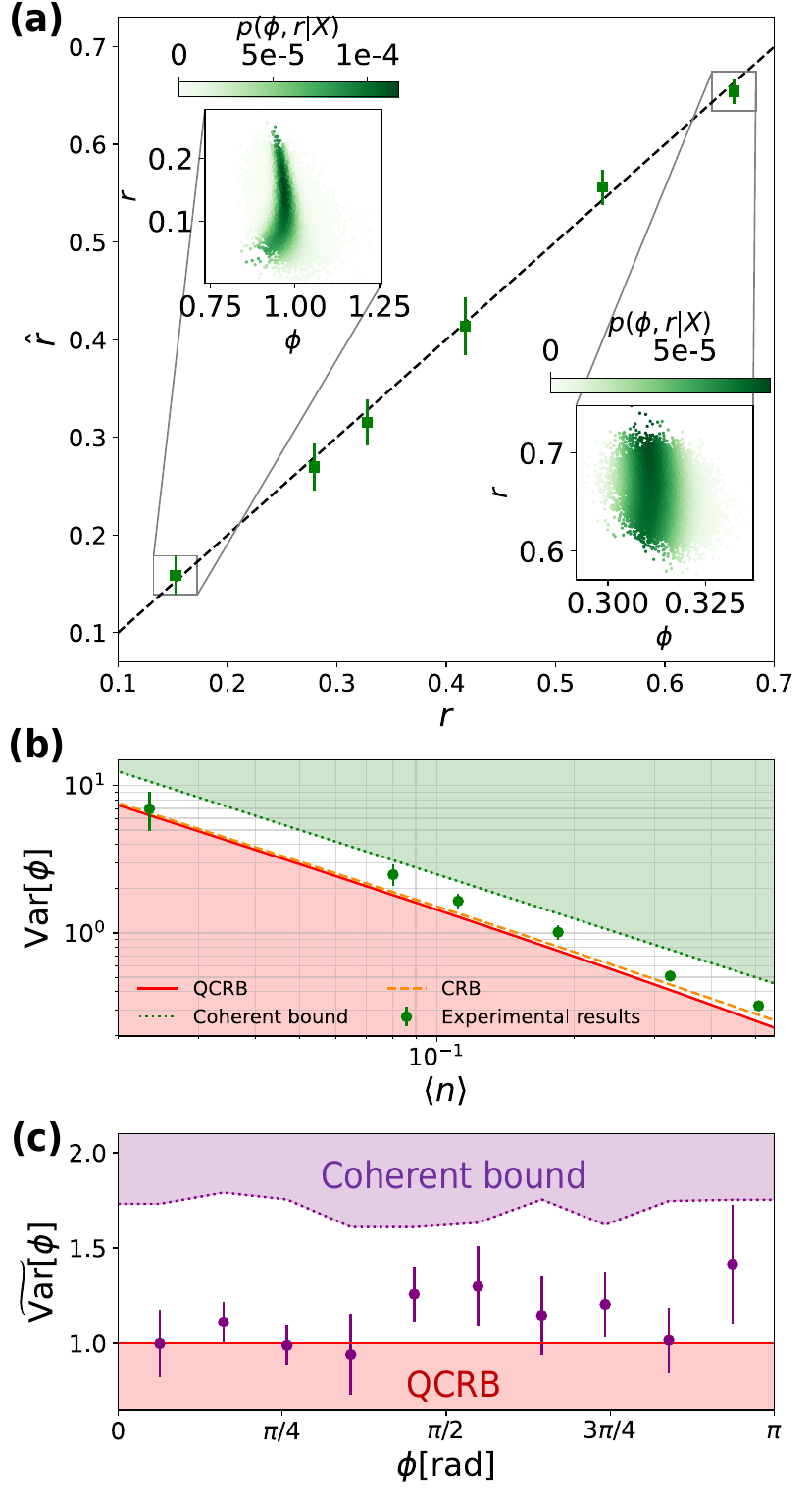}
\caption{\textbf{Robustness of the multiparameter adaptive estimation protocol to different squeezing strength and fully black-box approach.} In \textbf{(a)} the green squares represents the estimated squeezing level $\hat{r}$, within the multiparameter adaptive protocol, compared to the actual values $r$. As an example, we illustrate the experimentally reconstructed posterior distributions $p(\phi, r|X)$, corresponding to the lowest (top-left inset) and the highest (bottom right inset) squeezing levels. \textbf{(b)} Scaling of the experimental phase variances compared to the relative bounds: the QCRB for squeezed vacuum probes (red), the CRB (orange), and the QCRB for coherent states (green). The data lie all below the coherent bounds, computed without considering losses.
Panel \textbf{(c)}, reports the results obtained with a fully black-box approach, in which the efficiency $\eta$ is also estimated. The normalized variances $\widetilde{\mathrm{Var}}[\phi]$ (purple dots) are compared with the classical lossless (dash-dotted purple line) bound. In all the panels, the reported errors correspond to standard deviations over 10 repetitions of the experiments. }
\label{fig:s}
\end{figure}

A more direct comparison of the single and multiparameter strategies is reported in Fig.\ref{fig:adaptivity}, where we investigate how the estimation variances scale with the number of homodyne measurements. More specifically, in Fig.\ref{fig:adaptivity}a the experimental single-parameter posterior distribution is reported for a representative phase estimate as a function of the number of homodyne measurements used to update the Bayesian posterior. Fig.\ref{fig:adaptivity}b shows the reconstructed distribution at the end of the protocol, quantifying the achieved precision as $\mathrm{Std}[\phi] = \sqrt{\mathrm{Var}[\phi]}$.
Fig.\ref{fig:adaptivity}c illustrates the sequence of feedback values applied at successive steps of the estimation protocol. In the single-parameter case, after the rough estimate the feedback angles correspond to the quadrature maximizing the Fisher information, while in the multiparameter case, the rough estimation stage is split into three steps, $\theta = \{0, \pi/4, \pi/2\}$, to improve the estimation of $r$ needed to implement the adaptive feedback. The feedback angles for the fine estimate are instead determined by the condition where the off-diagonal elements of the FI matrix vanish (see Fig \ref{fig:adaptivity}d). 
The single-parameter strategy, which assumes prior knowledge of the squeezing parameter, required to set the feedback of the adaptive algorithm, displays faster convergence to the bound as emerges in Fig.\ref{fig:adaptivity}e. As expected, the multiparameter strategy requires a larger number of measurements before reaching convergence, since part of the data are used to estimate the squeezing level of the probe. Nevertheless, and most importantly, even in the absence of any prior calibration of the squeezing, the multiparameter protocol achieves phase estimation precision at sub-SQL level once convergence is reached. This demonstrates that quantum-enhanced performance can be retained even without relying on external information about the probe. Finally, the advantage over coherent-light phase estimation is further emphasized in the zoomed-in region showing the convergence behavior. Both our single-parameter (Fig.\ref{fig:adaptivity}f) and multiparameter (Fig.\ref{fig:adaptivity}g) strategies consistently outperform the lossless coherent bound calculated without rescaling by the overall transmission coefficient $\eta$.

We next demonstrate the full strength of the multiparameter protocol in a challenging regime where not only the optical phase, but also the squeezing parameter itself, undergoes significant variations, to test the robustness of this protocol to different squeezing levels. The estimated values of $r$ are shown in Fig.\ref{fig:s}a as a function of the calibrated squeezing.
Remarkably, the adopted strategy can faithfully infer the true value of $r$ across all operating conditions, without requiring any prior calibration or external reference. At the same time, the variance of the simultaneous phase estimate, reported for different squeezing regimes, remains consistently below the classical bound as shown in Fig.\ref{fig:s}b, confirming the ability of our protocol to preserve quantum-enhanced performance even under strongly different conditions of the probe. This is achieved since the adaptive feedback is explicitly designed to minimize the variance of the parameter of interest, namely the phase $\phi$. This feature is also related to the shape of the reconstructed posterior, which is narrower along the phase axis, while still providing reliable estimates of the squeezing. Our results highlight the versatility of the adopted multiparameter approach that enables the simultaneous estimation of probe and signal parameters, while allowing the feedback strategy to be tailored to prioritize the precision of the most relevant observable.
Moreover, by analyzing the dependence of the estimation variance on the squeezing strength $r$, which directly quantifies the photon number in the probe, we clearly observe an improvement with respect to the classical coherent bound. As $r$ increases, the variance decreases accordingly, and the gap between squeezed light estimation and the classical coherent state benchmark widens. This behavior provides an experimental indication that stronger squeezing yields greater metrological gain.

Finally, to make the protocol completely independent of any calibration, we tested its performance under the most demanding condition, in which no prior knowledge of the overall detection efficiency $\eta$ is also assumed. Estimating losses jointly with the phase is particularly challenging, since the corresponding parameters are incompatible observables \cite{ohno2025simultaneous}. To overcome this limitation, we adopt a hybrid strategy that combines Bayesian inference with a maximum-likelihood estimation of $\eta$. Specifically, during each update of the three-dimensional posterior, $\eta$ is set to the maximizer of the $\eta-$posterior at the previous step. 
This update bypasses full three-parameter inference, retaining the most consistent value of $\eta$ in every step of the estimation process without the need for additional measurement, focusing only on the physically relevant parameters $\phi$ and $r$.
This mixed approach allows us to maintain quantum-enhanced performance (see Fig.\ref{fig:s}c) without relying on any external calibration of the probe or the detection chain, thereby validating the fully unconditional nature of the protocol.

\parTitle{Discussion/Conclusion}
Our work presents the first experimental demonstration of adaptive multiparameter quantum phase estimation with squeezed vacuum states, achieving sub-SQL precision across the full periodicity interval $[0,\pi)$ without relying on prior calibration of key system parameters such as the squeezing level and detection efficiency. These parameters are known to fluctuate in realistic scenarios, both due to temporal drifts of the source and to sample-dependent transmission, making calibration-based approaches inherently unstable and ultimately compromising the effectiveness of adaptive measurement strategies. 
By implementing an adaptive multiparameter protocol that simultaneously infers the phase $\phi$ and the relevant nuisance parameters, we overcome this calibration bottleneck while retaining unconditional quantum advantage over the entire unambiguous phase range.
Crucially, the method is robust to consistent fluctuations of $r$, maintaining quantum-enhanced performance across a wide operating range. 

These findings demonstrate the feasibility of self-calibrating, ab-initio quantum metrology protocols that are resilient to probe instabilities and experimental drifts, thereby extending the practical applicability of squeezed light for precision sensing. Beyond enabling reliable operation under realistic conditions, our results establish a pathway toward scalable implementations in advanced interferometric platforms and distributed quantum sensor networks, which rely on the simultaneous estimation of multiple parameters.

\section*{Acknowledgments}

This work is supported by the Amaldi Research Center funded by the  Ministero dell'Istruzione dell'Università e della Ricerca (Ministry of Education, University and Research) program ``Dipartimento di Eccellenza'' (CUP:B81I18001170001), by the project QU-DICE, Fare Ricerca in Italia, Ministero dell'istruzione e del merito, code: R20TRHTSPA, and by the PNRR MUR project PE0000023-NQSTI (Spoke 4). We thank M. Barbieri for the useful discussion and the Noisy Labs team.

\bibliographystyle{naturemag}
\bibliography{biblio}

\end{document}

\beginsupplement

\title{\textit{Supplementary Information for}:\\ Multiparameter quantum-enhanced adaptive metrology with squeezed light}

\author{Giorgio Minati}
\affiliation{Dipartimento di Fisica, Sapienza Universit\`{a} di Roma, Piazzale Aldo Moro 5, I-00185 Roma, Italy}

\author{Enrico Urbani}
\affiliation{Dipartimento di Fisica, Sapienza Universit\`{a} di Roma, Piazzale Aldo Moro 5, I-00185 Roma, Italy}

\author{Nicolò Spagnolo}
\affiliation{Dipartimento di Fisica, Sapienza Universit\`{a} di Roma, Piazzale Aldo Moro 5, I-00185 Roma, Italy}

\author{Valeria Cimini}
\email{valeria.cimini@uniroma1.it}
\affiliation{Dipartimento di Fisica, Sapienza Universit\`{a} di Roma, Piazzale Aldo Moro 5, I-00185 Roma, Italy}

\author{Fabio Sciarrino}
\affiliation{Dipartimento di Fisica, Sapienza Universit\`{a} di Roma, Piazzale Aldo Moro 5, I-00185 Roma, Italy}

\maketitle

\section{Experimental Setup Characterization}
\subsection{Experimental Setup}
\begin{figure*}[htb!]
\centering
\includegraphics[width=0.8\textwidth]{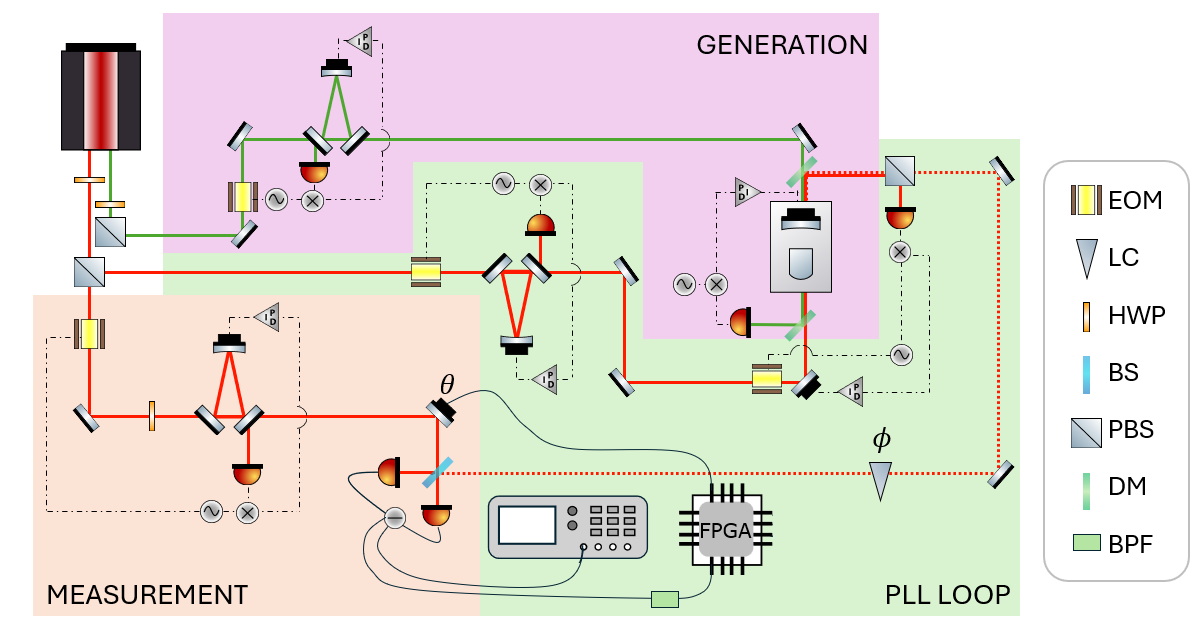}
\caption{\textbf{Experimental setup.} The laser provides two outputs. The $532$ nm beam pumps the squeezer cavity, while the $1064$ nm beam is split by a half-wave plate (HWP) and a polarizing beam splitter (PBS) into a control beam and a local oscillator (LO), both mode-matched to the squeezed output.
The setup comprises three main blocks: (i) squeezed-state generation, (ii) homodyne measurement, and (iii) phase stabilization using the control beam in a phase-locked loop (PLL). 
The squeezer cavity is stabilized via a Pound–Drever–Hall (PDH) scheme employing an electro-optic modulator (EOM), a piezoelectric actuator, and a photodiode; the same infrastructure is used to keep three mode-cleaner cavities resonant with their respective beams. After the squeezer, the $532$ nm pump and the $1064$ nm control/squeezed beams are separated with dichroic mirrors (DM). The phase $\phi$ along the squeezed-beam path is tuned with a liquid-crystal (LC) phase shifter and measured by balanced homodyne detection (HD), where the LO and the squeezed beam interfere on a 50:50 beam splitter (BS). The quadrature angle $\theta$ is swept using a piezoelectric stage driven by an FPGA, which reads the DC output of the homodyne detector after a band pass filter (BPF) to suppress electronic noise. } 
\label{fig:setup}
\end{figure*}

In what follows, we present a complete description of the setup employed to perform the experiment, which is schematically illustrated in Fig.\ref{fig:setup}. 

The main laser source is a Coherent Prometheus Nd:YAG double-wavelength continuous-wave laser emitting 1.77 W at 1064 nm and 104 mW at 532 nm. The 532 output serves as the pump for the squeezed state generation. Before this step, the action of a Half-Wave Plate (HWP) and a Polarizing Beam Splitter (PBS) allows us to tune the total power of the 532 nm beam. Phase modulation in the MHz frequency range is applied using an electo-optical modulator (EOM) to produce sidebands used for the Pound-Drever-Hall (PDH) locking.
%\cite{drever1983laser, black2001introduction}. 
All the beams employed are cleaned through three mirrors Fabry-Perot ring cavities, each one used as an optical resonator for mode-cleaning (MC), filtering out spatial modes different from the $\mathrm{TEM}_{00}$.

\emph{Generation stage.} The squeezed light source is the commercial source developed by Noisy Labs. It consists of an optical parametric amplifier based on a type-0 periodically poled KTP (PPKTP) crystal configured as a hemilithic standing-wave cavity pumped via the mode-cleaned $532$ nm beam. One mirror is piezo-mounted for length control, while the high-reflection coating on one end face of the PPKTP acts as the second mirror at 1064 nm (and appropriately coated for 532 nm). The opposite crystal face is anti-reflection coated. The cavity is locked on resonance via PDH. The crystal temperature is actively stabilized to maintain quasi-phase matching at the pump wavelength and to suppress slow thermal drifts; length control is handled by the piezo and fine-tuned with an additional temperature control. The output of the source can be extracted using a Dicroic Mirror (DM) that reflects only 1064 nm wavelength.
The 1064 nm beam is split into two paths, each subjected to the same mode-cleaning procedure used for the pump. This ensures high spatial purity and excellent overlap at the detection stage between the squeezed field and the local oscillator (LO), which is essential for resolving large levels of squeezing. One path forms the LO for balanced homodyne detection, while the second serves as a coherent control beam used to stabilize the phase of the squeezed field.\\
\emph{Measurement stage.} We employ a balanced homodyne detection scheme where the LO and the squeezed beam are mixed on a 50:50 beam splitter (BS) and the photocurrent difference is measured via an oscilloscope. The detected quadrature is set by the LO phase $\theta$ controlled via a motorized piezoelectric stage.\\
\emph{Phase locking loop.} The phase of the squeezed beam is stabilized using a coherent control beam that is spatially mode-matched to the squeezed mode. The control beam is phase-modulated by an EOM at radio frequencies and injected in the squeezer cavity. When in resonance, the bright leakage provides an alignment reference for the squeezed path; at low power it generates a phase-sensitive error signal upon interference with the LO. This signal is demodulated and used in a phase-locked loop (PLL).

In this configuration, the OPA produces squeezed vacuum at 1064 nm the MHz detection band. The combination of mode cleaning, PDH-stabilized cavities, and coherent-control locking provides the spatial mode quality and phase stability $\theta-\phi$ required to resolve sub-shot-noise quadrature fluctuations and to implement the adaptive, real-time estimation protocol.

\subsection{Squeezed light characterization}

\begin{figure*}[htb!]
\centering
\includegraphics[width=0.5\textwidth]{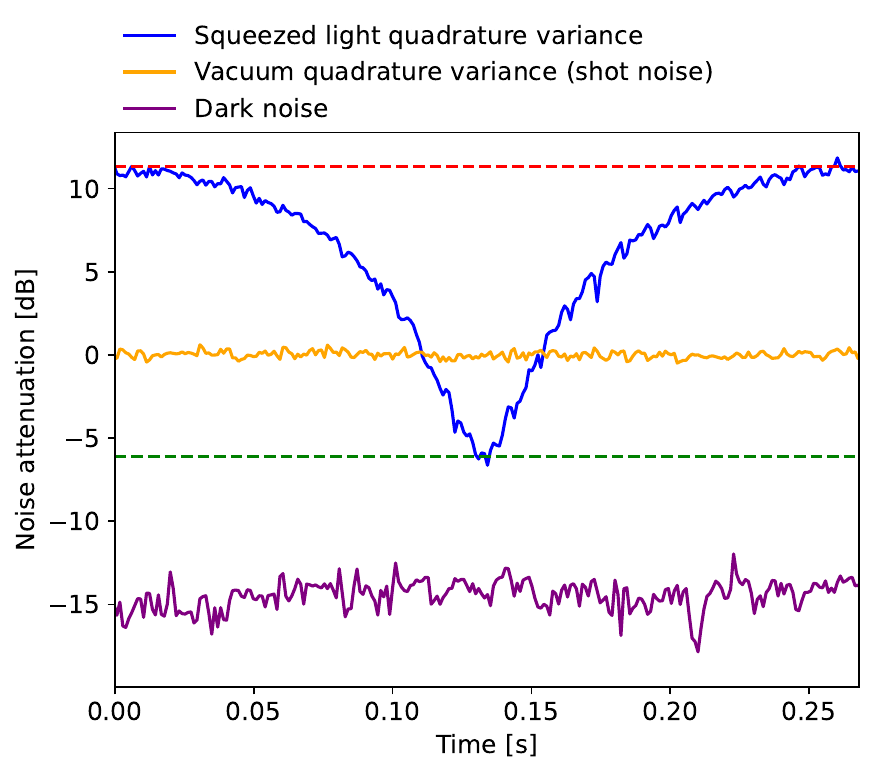}
\caption{\textbf{Squeezed quadratures noise attenuation, compared to the shot noise and the dark noise.} In this figure, we report the squeezed light variance attenuation (blue trace), in units of shot-noise. In particular, it has been acquired by scanning the LO phase by ramping the piezo mirror with a triangular wave with a frequency of 1 Hz. Additionally, we also report the shot noise (yellow trace) and the electronic dark noise (purple trace). All three traces have been acquired with a zero-span measurement at 6 MHz.
}
\label{fig:sqz_level}
\end{figure*}

In order to characterize the generated squeezed state, we measure the photocurrent noise modulation in different configurations, which we report in Fig.\ref{fig:sqz_level}, where all the traces are zero-span measurements at 6 MHz. We take the shot-noise as a reference, hence, setting the noise level which is measured when only the LO is sent to the homodyne detector at 0 dB. A 1 Hz triangular-wave-shaped scan of the measured quadrature angle allows us to estimate the maximum and minimum noise modulation (blue trace in Fig.\ref{fig:sqz_level}), thereby quantifying the squeezing and anti-squeezing level as $\sigma_{\mathrm{sqz}}^2 = -6.12 \pm 0.28 \ \mathrm{dB}$ and $\sigma_{\mathrm{asqz}}^2 = 11.35 \pm 0.28 \ \mathrm{dB}$. Additionally, within the same settings, we measure the electronic dark-noise when no light at all is sent to the detector, obtaining a noise attenuation of $\sigma_{\mathrm{dark}}^2=  -14.9 \pm 1.0 \ \mathrm{dB}$

\subsection{Phase Locking Stability and Control}
\label{subsec:phase_lock}
In what follows, we describe the implementation of the FPGA-based locking procedure employed to measure the squeezed field quadrature along an arbitrary angle.
First, we model the fields incoming at the homodyne detectors as:
\begin{align}
    E_1 &= E_{\mathrm{LO}} e^{i (\omega t + \vartheta )},\\
    E_2 &= E_{\mathrm{CF}} e^{i (\omega t + \beta \sin{(\Omega t)})}.
\end{align}
The first is the LO field, oscillating with an amplitude $E_{\mathrm{LO}}$ at the optical frequency $\omega$, with a phase $\vartheta$ set by the LO piezo mirror. The second field is the weak coherent field superimposed onto the squeezed light path, whose oscillation has amplitude $E_{\mathrm{CF}}$ and, in addition to the optical frequency $\omega$, is phase-modulated with sidebands of frequency $\Omega$ and modulation depth $\beta$. The interference of these in the HD detection BS produces the output fields $E_3$ and $E_4$, which are computed as follows:
\begin{equation}
    \begin{pmatrix}
        E_3\\
        E_4
    \end{pmatrix} = \frac{1}{\sqrt{2}}
    \begin{pmatrix}
        1 &i \\
        i & 1
    \end{pmatrix}
    \begin{pmatrix}
        E_1\\
        E_2
    \end{pmatrix} = \frac{1}{\sqrt{2}}
    \begin{pmatrix}
        E_1 + i E_2\\
        i E_1 + E_2
    \end{pmatrix}.
\end{equation}
The corresponding photocurrents measured at the HD detector are proportional to the output fields intensities $P_3$ and $P_4$:
\begin{align}
    P_3 & = |E_3|^2 = \frac{1}{2} (|E_1|^2 + |E_2|^2 + i E_1^* E_1 - E_1 E_1^*),\\
    P_4 & =  |E_4|^2 = \frac{1}{2} (|E_1|^2 + |E_2|^2 - i E_1^* E_1 + E_1 E_1^*), 
\end{align}
therefore, the photocurrents subtraction reads:

\begin{align}
    P_{\mathrm{HD}} = P_3 - P_4 & = i (E_1^* E_2 - E_1 E_2^*) =\\
            & = i E_{\mathrm{LO}} E_{\mathrm{CF}} 
                \left( 
                e^{i(-\vartheta + \beta \sin{(\Omega t)})} -
                e^{-i(-\vartheta + \beta \sin{(\Omega t)})}
                \right) =\\ \label{eq:bessel}
            & \simeq i E_{\mathrm{LO}} E_{\mathrm{CF}} 
                \left( 
                e^{-i\vartheta} \left[ J_0(\beta) + 2 i J_1(\beta) \sin{(\Omega t)} \right] -
                e^{i\vartheta} \left[ J_0(\beta) - 2 i J_1(\beta) \sin{(\Omega t)} \right]
                \right) =\\ \label{eq:HD_signal}
            & = 2 E_{\mathrm{LO}} E_{\mathrm{CF}} \left( J_0(\beta) \sin{(\vartheta)} - 2 J_1(\beta) \sin{(\Omega t)} \cos{(\vartheta)} \right),
\end{align}
where, in \eqref{eq:bessel}, the terms $e^{\pm i\beta \sin{(\Omega t)}}$ have been expressed as a Bessel series truncated at order 1.
The part of Eq.\eqref{eq:HD_signal} oscillating at $2 \Omega$, selected using a high-pass filter, is then mixed with an electronic local oscillator $E_3 = E_{\mathrm{eLO}} \sin{(\Omega t+ \varphi_{\mathrm{demod}})}$, obtaining;
\begin{align}
     P_{\mathrm{HD}}^{\mathrm{mix}} &= - 4 E_{\mathrm{LO}} E_{\mathrm{CF}} J_1(\beta) \sin{(\Omega t)} \cos{(\vartheta)} \ \cdot \  E_{\mathrm{eLO}} \sin{(\Omega t+ \varphi_{\mathrm{demod}})} = \\
     &= - 4  E_{\mathrm{LO}} E_{\mathrm{CF}} E_{\mathrm{eLO}} J_1(\beta) \cos{(\vartheta)} \left( \sin{(2 \Omega t + \varphi_{\mathrm{demod}}) + \sin{(\varphi_{\mathrm{demod}})}}  \right).
\end{align}
A low-pass filter eliminates the terms oscillating at $2 \Omega$, resulting in the following error signal:
\begin{equation}
\label{eq:errcos}
    P_{\mathrm{cos}}^{\mathrm{err}} = - 4  E_{\mathrm{LO}} E_{\mathrm{CF}} E_{\mathrm{eLO}} J_1(\beta) \cos{(\vartheta)} \sin{(\varphi_{\mathrm{demod}})}.
\end{equation}

Therefore, we have access to the first component of \eqref{eq:HD_signal} (which we will call $P_{\mathrm{sin}}^{\mathrm{err}}$ in what follows) and \eqref{eq:errcos}, respectively modulated as $\sin{(\vartheta)}$ and $\cos{(\vartheta)}$. 
We can use them to generate an error signal that can lock the homodyne measure to an arbitrary angle $\alpha$. In fact, upon rescaling $P_{\mathrm{cos}}^{\mathrm{err}}$ by a factor $S$ to match the amplitude of $P_{\mathrm{sin}}^{\mathrm{err}}$, we can combine them in a sum with weights $A_{\alpha}, B_{\alpha}$ such that $A_{\alpha}^2 + B_{\alpha}^2 =1$. Hence, the final error signal will be:
\begin{equation}
    P_{\mathrm{err}}^{\alpha} = A_{\alpha} P_{\mathrm{sin}}^{\mathrm{err}} + B_{\alpha} SP_{\mathrm{cos}}^{\mathrm{err}} \sim \sin(\vartheta + \alpha),
\end{equation}
where $\alpha = \arctan{(B_{\alpha}/A_{\alpha})}= \arccos{(A_{\alpha})} = \arcsin{(B_{\alpha})}$. In what follows, for notation simplicity, we define $\theta = \vartheta + \alpha$.

\begin{figure*}[htb!]
\centering
\includegraphics[width=0.9\textwidth]{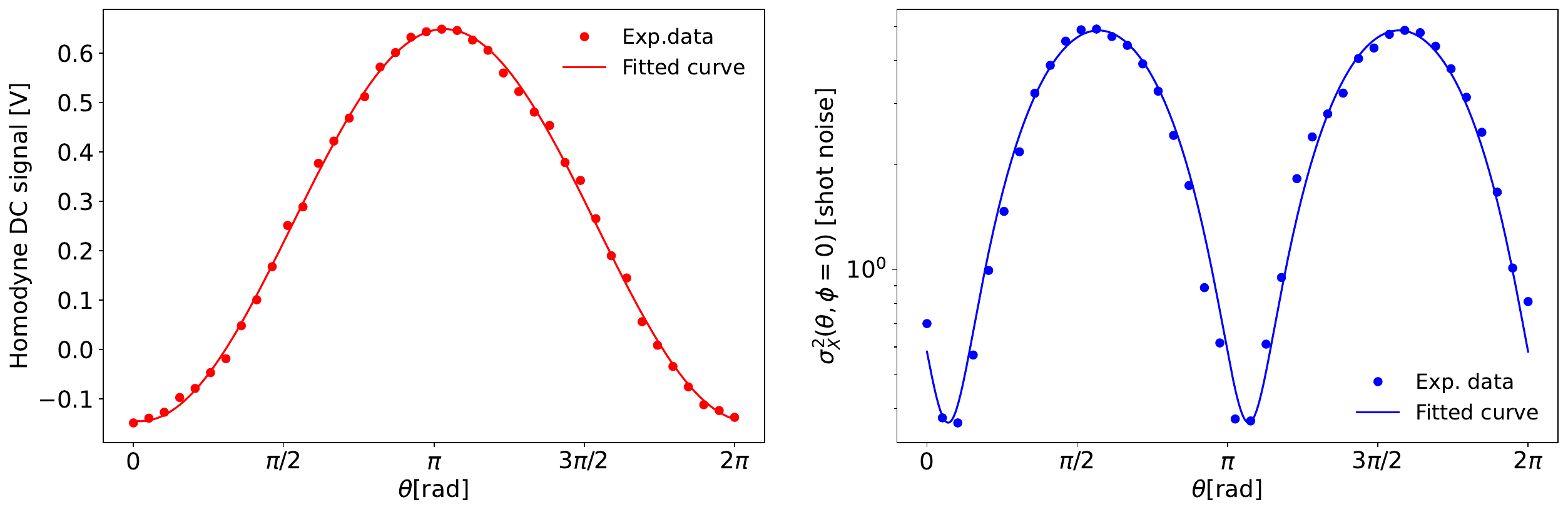}
\caption{\textbf{Phase-locked measurements.} In panels \textbf{(a)} and \textbf{(b)} are reported, the homodyne DC signal (red dots) and the quadratures variance (blue dots) acquired for different phase-locked homodyne angles spanning the range $\theta \in [0,2 \pi]$, respectively. Both experimental datasets have been acquired by averaging over $200,000$ acquisitions, but in the second case, a band-pass filter around $1.1$ MHz is also required, especially to remove low-frequency noise.
We report the corresponding fitted curves, as red and blue solid lines for the homodyne DC and quadratures variance, respectively. The former employs a cosine model, while the latter uses the lossy variance model for the quadrature variance defined in Eq.(4) of the main text.
}
\label{fig:stabilità}
\end{figure*}

To experimentally test this procedure, we lock the homodyne measure to different $\theta \in [0 ,2 \pi]$ and, for each of them, we measure the resulting homodyne DC signal and the quadrature variance, and report the corresponding results in the left and right panels of Fig.\ref{fig:stabilità}, respectively. The regularity of these measurements demonstrates the reliability of this approach, and enables, by fitting the experimental data, to finely calibrate possible overall offsets of the phase $\theta$.

\section{Additional Experimental Data}

\begin{figure*}[htb!]
\centering
\includegraphics[width=0.99\textwidth]{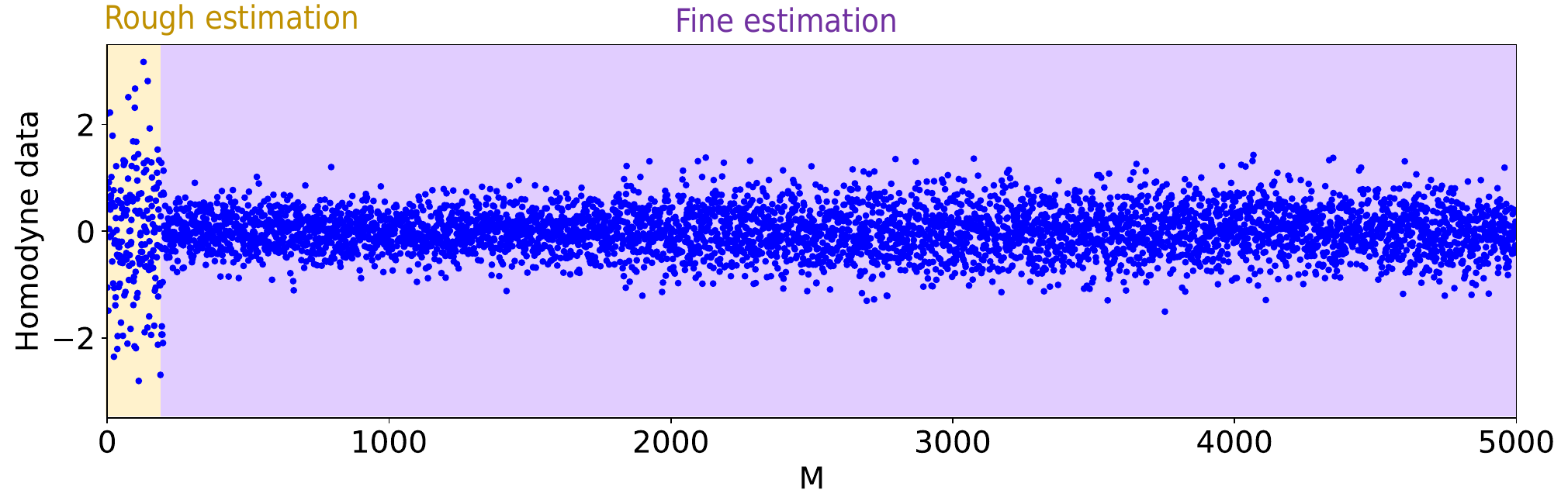}
\caption{\textbf{Homodyne data during the adaptive protocol.} Here we plot the homodyne measurements taken during the single-parameter estimation of the phase $\phi = 1.1$ rad. The normalized measurements are reported in such a way to fix the shot-noise to the conventional value $\sigma_{\mathrm{shot}}^2 = 1/4$. The yellow background indicates the fraction of measurements employed to obtain the phase rough estimation necessary to compute the feedback for the LO phase. Its effect is noticeable in the second stage of the measurement, illustrated on a purple background, where the homodyne data exhibits a smaller variance, corresponding to measuring along the optimal quadrature angle.
}
\label{fig:X_feedback}
\end{figure*}

In this Section, we describe additional measures relative to the results reported in the main text. To better understand the effects of adaptive feedback, we report in Fig.\ref{fig:X_feedback} the quadratures measured for a specific phase $\phi$ during the estimation protocol. We can observe how a small portion of the total acquired data (illustrated with a yellow background in Fig.\ref{fig:X_feedback}) is sufficient to have a rough estimation, precise enough to compute a feedback that shifts the LO phase towards the optimal one, noticeable by the substantial decreasing of the quadrature variance in the second stage of the estimation (depicted in Fig.\ref{fig:X_feedback} with a purple background).

\begin{figure*}[htb!]
\centering
\includegraphics[width=0.99\textwidth]{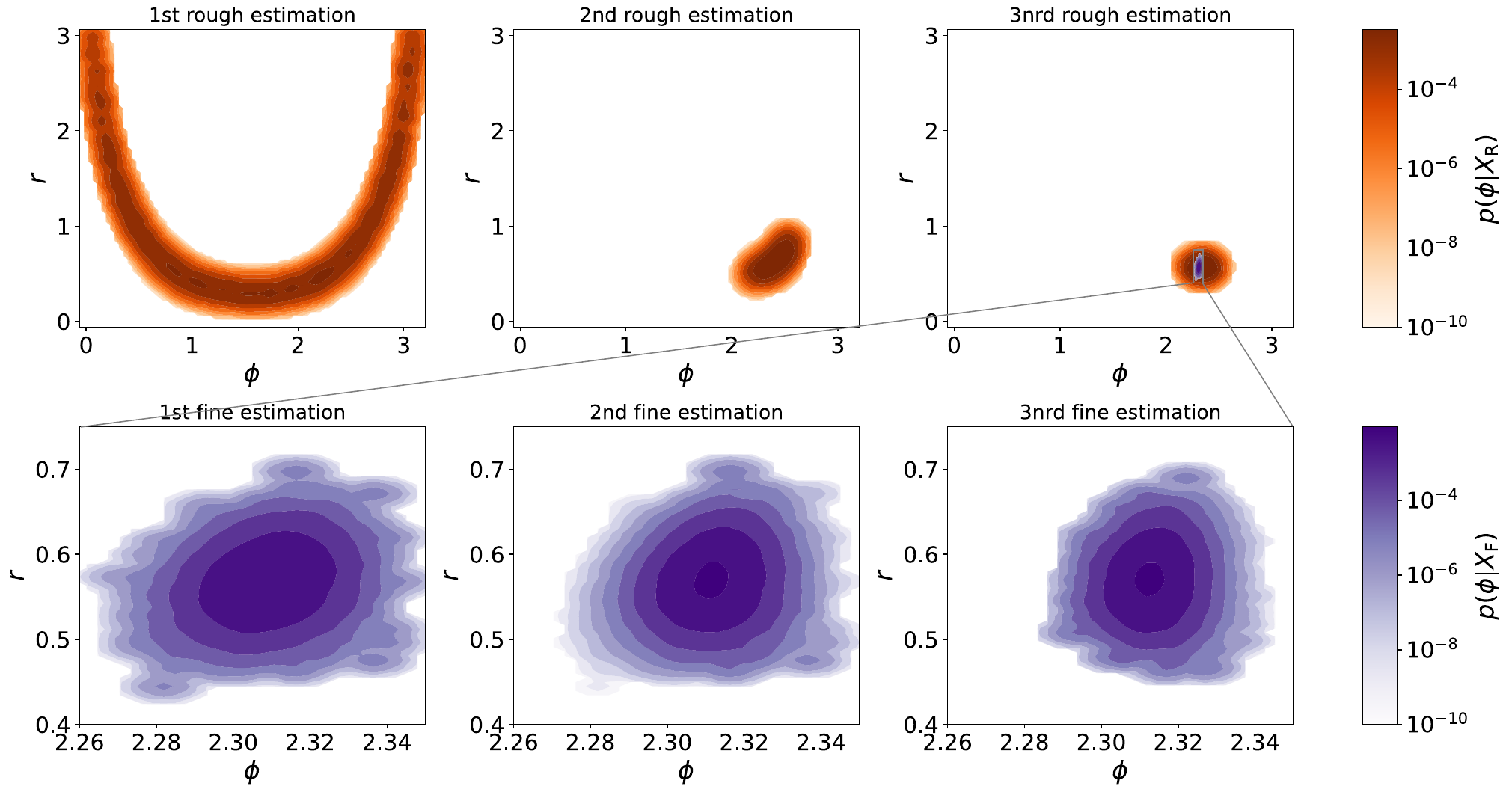}
\caption{\textbf{Evolution of the experimental posterior distribution during the multiparameter adaptive protocol.} In this figure, the posterior distribution $p(\phi, r|X)$ is updated during the different steps of the adaptive protocol. In the upper panels, we report the posterior during the three steps of the rough estimation, using an orange color scale. Instead, the three bottom panels illustrate the posterior evolution during the fine estimate (purple color scale), again subdivided into three adaptive feedback steps. }
\label{fig:posterior_2par}
\end{figure*}

In the case of multiparameter estimation, we can gain further insights into the functioning of the adaptive estimation by studying how the experimental 2-dimensional posterior $p(\phi, r|X)$ evolves during the different steps of the protocol, illustrated in Fig.\ref{fig:posterior_2par}. After the first step of the rough estimation (upper left panel in Fig.\ref{fig:posterior_2par}), taken with the LO phase set at $\theta=0$, the posterior presents a symmetric behavior along the phase. This feature reflects the fact that a homodyne data taken in a single setting cannot disambiguate the phase between a phase $\phi$ and $\pi - \phi$. The second step of the protocol (upper central panel in Fig.\ref{fig:posterior_2par}) updates the previous posterior with quadratures measured along $\theta=\pi/4$, and illustrates how this additional measurement setting resolves the ambiguity mentioned above. The third part of the rough estimation, by setting $\theta=\pi/2$, probes the orthogonal quadrature with respect to the first step, improving, as we can see in the upper right panel of Fig.\ref{fig:posterior_2par},  the estimation of the squeezing parameter. A sufficiently precise estimate of $r$ is essential for computing accurate adaptive feedback, whose effect becomes clear when comparing the final rough estimate with the initial fine estimation (purple color scale). The lower panels show the posterior distributions for the three steps of the fine estimation on a magnified scale, illustrating how much of the overall precision gain is attributable to the adaptive feedback.

\begin{figure*}[htb!]
\centering
\includegraphics[width=0.99\textwidth]{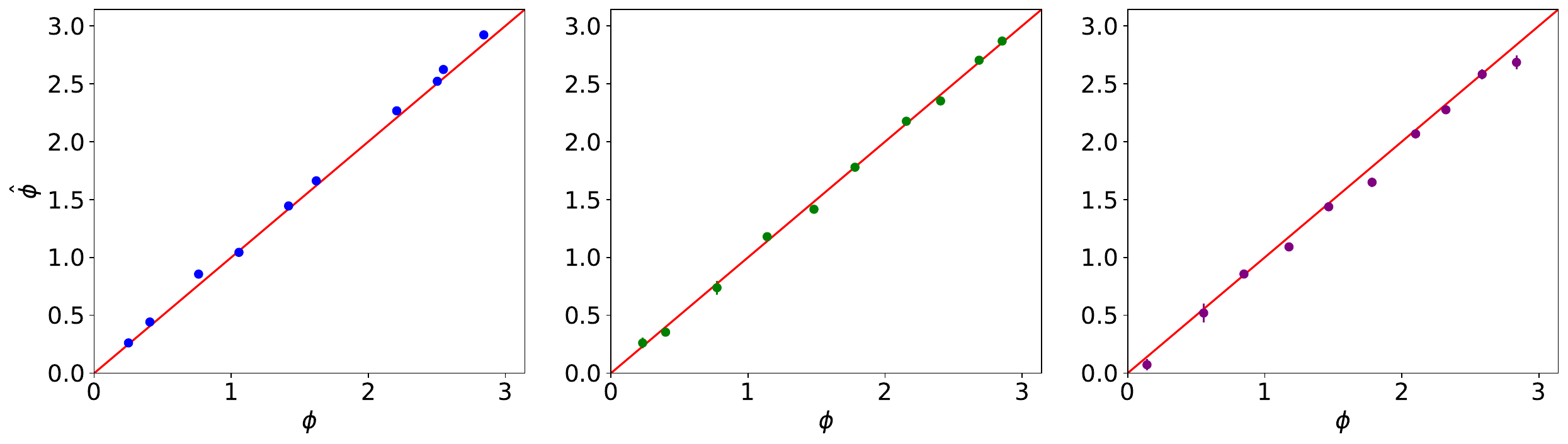}
\caption{\textbf{Estimated phases compared to the calibrated phase values.}
In the three panels, we report the estimated phase values $\hat{\phi}$ as a function of the actual phase $\phi$, respectively for the single parameter (left panel, blue dots), 2-parameter (central panel, green dots), and 3-parameter (right panel, purple dots) protocols.}
\label{fig:est_vs_true}
\end{figure*}

To quantify the protocol ability to estimate the phase ab-initio over the full $[0,\pi)$ range, the estimated values of the $10$ different investigated phases $\hat{\phi}$ are compared in Fig.\ref{fig:est_vs_true} to the true calibrated values $\phi$ for all the inspected strategies. Starting from the experimental single-parameter estimate, the estimated values of the two-parameter ones, and the black box variant of the protocol that also infers the overall detection efficiency $\eta$ from the data, are here reported.  

The scaling of the phase variance for the black-box protocol, which jointly estimates $(\phi, r, \eta)$, as a function of the number of measurements is reported in Fig.\ref{fig:scaling_n}. Even in this case, the implemented approach enables a phase estimate below the coherent bound, confirming that the information gain provided by squeezing is retained even while the squeezing parameter and the overall efficiency are inferred on the fly and not pre-calibrated.

\begin{figure*}[htb!]
\centering
\includegraphics[width=0.85\textwidth]{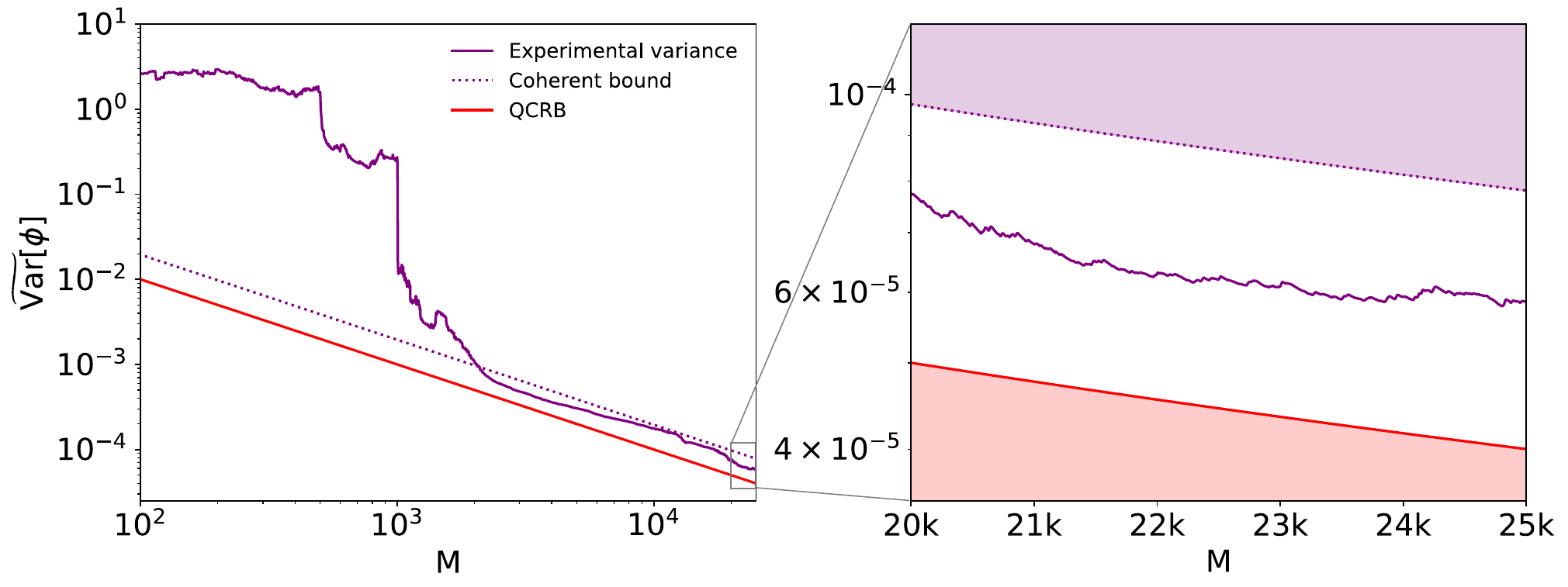}
\caption{\textbf{Scaling of the phase estimation variance as a function of the number of measurements, in the case of the 3-parameter estimation protocol.} The normalized variance $\widetilde{\mathrm{Var}}[\phi] \equiv \mathrm{Var}[\phi]\cdot M F_{Q}^{\mathrm{sq}}$ is reported as a function of the homodyne measurements $M$ and compared to the QCRB (solid red line). The dotted line represents the coherent bound for the lossless scenario.
}
\label{fig:scaling_n}
\end{figure*}

In terms of scaling with the average number of photons in the probe, in the ideal lossless case, the variance of the phase estimate retrieved with coherent probes scales at the SQL, $\mathrm{Var}[\phi]\sim \frac{1}{4\langle n\rangle}$. On the contrary, squeezed probes attain a quantum-enhanced scaling $\frac{1}{8\langle n \rangle (\langle n \rangle +1)}$ that for $\langle n \rangle>>1$ attains the Heisenberg limit.
Such behaviors are altered in the presence of losses, therefore, when $\eta<1$ the asymptotes change, obtaining bounds with slower power laws as emerges in the plot in Fig.\ref{fig:scaling_n}.  However, even for realistic lossy conditions, squeezed probes retain an advantage over coherent probes, and this is the region where we work experimentally.

\begin{figure*}[htb!]
\centering
\includegraphics[width=0.6\textwidth]{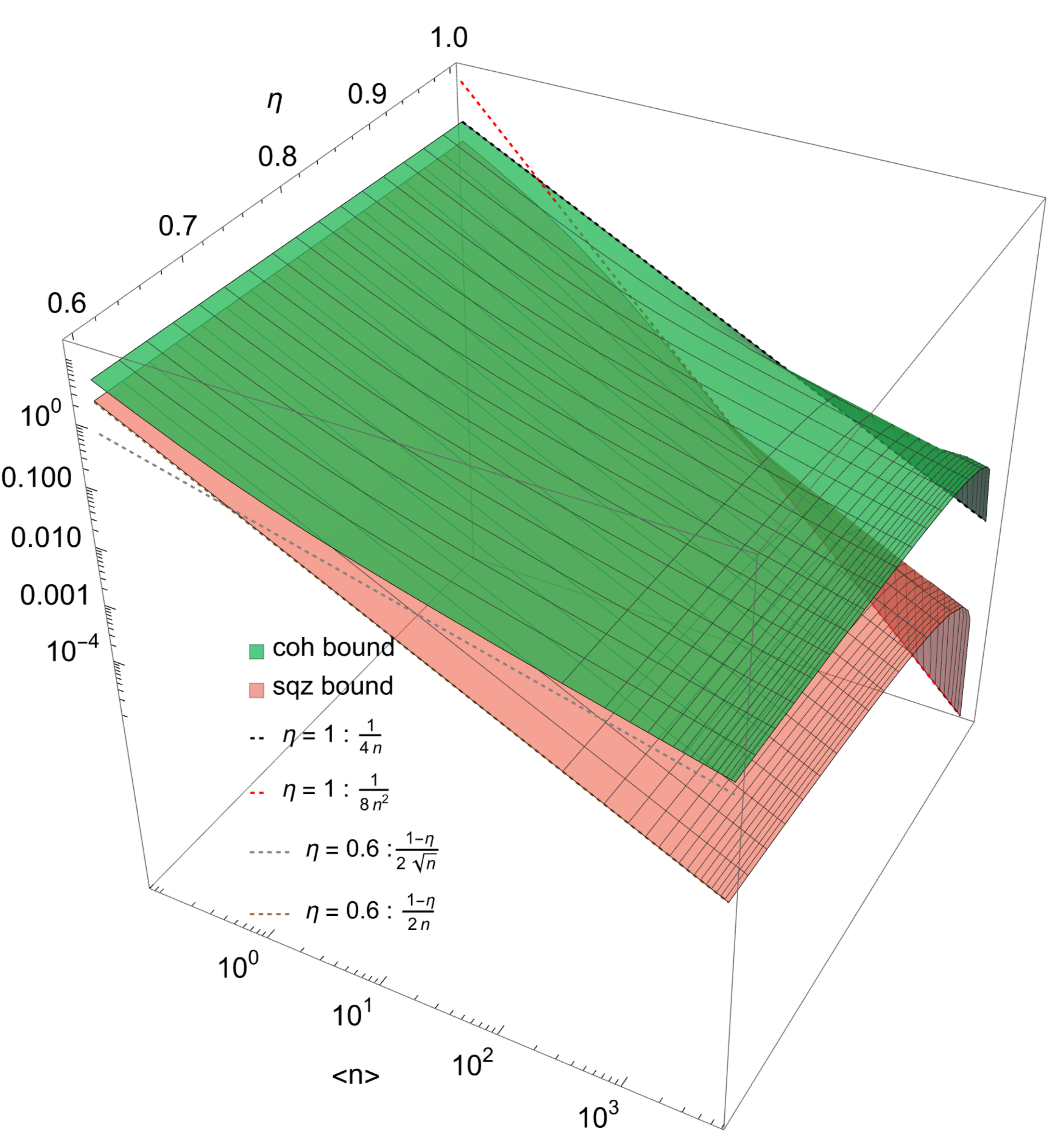}
\caption{\textbf{Scaling of the phase estimation variance as a function of the probe average photon number.} The ultimate phase–variance bounds for coherent (green surface) and squeezed (red surface) probes as functions of the mean photon number $\langle n \rangle$ and total efficiency $\eta$ are reported. In the lossless limit ($\eta=1$), coherent light follows the SQL, while squeezed probes achieve a quadratic scaling. Loss degrades these scalings. 
}
\label{fig:scaling_n}
\end{figure*}

\section{Bounds on simultaneous estimation of phase and squeezing level}

\subsection{Quantum Fisher Information and Compatibility Conditions}
\label{sec:QFI}

In order to find the ultimate precision bound for the estimation of multiple parameters $\Vec{y} = \{y_i\}_{i=1}^P$ from a given quantum state $\rho$, we have to compute the corresponding Quantum Fisher Information (QFI) matrix $\bm{F_Q}$. It defines the inequality chain reported in Eq.(4) in the main text, which, for the sake of clarity, we also rewrite here:
\begin{equation}
\label{eq:multi_CRB}
    \bm{\Sigma}[\Vec{y}] \overset{\mathrm{CRB}}{\succeq} \frac{1}{M} \bm{F}^{-1}[\{y_i\}_{i=1}^P] \overset{\mathrm{QCRB}}{\succeq} \frac{1}{M} \bm{F_Q}^{-1}.  
\end{equation}
In general, the elements of the QFI matrix can be computed as:
\begin{equation}
    (\bm{F_Q}[\rho])_{ij} = \frac{1}{2}\Tr \left(\rho \{L_i, L_j\} \right),
\end{equation}
where the curly brackets denote the anticommutator and $L_i$ is the so-called Symmetric Logarithmic (SLD) operator corresponding to the estimation of $y_i$. This operator is indirectly defined by the equation:
\begin{equation}
    \frac{\partial \rho}{\partial y_i} = \frac{1}{2} \left( L_i \rho + \rho L_i \right),
\end{equation}
and plays a crucial role in quantum metrology, since not only it is needed for the computation of the QFI, but it also defines the possibility of actually saturating the QCRB when $\Tr[\rho[L_i,L_j]]=0$.
Let us consider parameters that are encoded in a pure state $\rho= \vert \psi_{\vec{y}} \rangle \langle \psi_{\vec{y}} \vert$ by means of unitary transformations $\vert \psi_{\vec{y}} \rangle = U(\vec{y}) \ket{\psi_0} = e^{-i\sum_i y_i G_i} \ket{\psi_0}$, where $G_i$ is the generator of the $i$-th transformation. The computation of the SLDs and the QFI is simplified as follows:
\begin{align}
\label{eq:SLD_G}
    L_i &= -2 i [G_i, \rho]\\
    (\bm{F_Q}[\vert \psi_{\vec{y}} \rangle])_{ij} &= 4 \left(\expval{G_i G_j} - \expval{G_i}\expval{G_j} \right),
    \label{eq:QFI_G}
\end{align}
where the expectation values are taken over the state $\vert \psi_{\vec{y}} \rangle$.\\
In our case, we consider a model in which the parameters of interest describe a pure squeezed vacuum state, featuring a squeezing level $r>0$ and rotated by a phase $\phi \in [0,\pi]$. The encoding of both these parameters can be modeled as the following unitary transformation of the vacuum state $\ket{0}$:
\begin{equation}
\label{eq:sqz_state}
    \ket{\phi,r} = e^{i (\phi G_{\phi} + r G_r)} \ket{0}, 
    \quad \mathrm{where} \quad G_{\phi}=n=a^{\dag}a, 
    \ G_r = \frac{i}{2}\left( a^{\dag 2} - a^2 \right).
\end{equation}
Then, the QFI corresponding to the operations reported in Eq.\eqref{eq:sqz_state} can be computed by means of Eq.\eqref{eq:QFI_G}, where we substitute $\Vec{y} = \{\phi, r\}$, obtaining the following matrix:
\begin{equation}
\label{eq:QFI_sqz}
    \bm{F_Q}[\ket{\phi, r}] = 
    \begin{pmatrix}
        2 \sinh^2{(2r)} & 0 \\
        0 & 2
    \end{pmatrix}.
\end{equation}
The diagonal form of the QFI in Eq.\eqref{eq:QFI_sqz} further simplifies the evaluation of ultimate precision bounds on the estimation of $\phi$ and $r$, which ultimately reduces to the inverse of the corresponding diagonal elements.\\
In a multiparameter setting, two parameters are said to be compatible if there exists a measurement strategy that can simultaneously saturate the QCRB for both parameters. A sufficient condition to achieve compatibility is the commutation of the SLDs corresponding to the parameters of interest, as projective measurements onto their common eigenbasis enable the optimal estimation of such parameters. When the state $\rho$ is pure, as in our model, a weaker necessary and sufficient condition can be verified, i.e. the so-called \textit{weak commutation relation}, which is verified $\expval{[L_i,L_j]}=0$. In our model, where the parameters are encoded via unitary transformation this expression further simplifies, and can be evaluated as follows:

\begin{equation}
\label{eq:weak_comm}
    \expval{[L_{\phi},L_r]} = 4 \expval{[G_{\phi},G_r]} = -4 i \sinh{(2r)} \neq 0.
\end{equation}
This means the parameters $\phi$ and $r$ are incompatible, i.e. we cannot devise a measurement strategy that simultaneously achieves the optimal precision in both estimations.\\
In what follows, we describe how adaptivity can exploit the multiparameter estimation framework to approach the phase estimation QCRB when the squeezing level is unknown.

\subsection{Adaptive protocol model}

The likelihood  $p(x|\theta, \phi, r)$ for a homodyne measurement with local-oscillator phase $\theta$ on a squeezed state (phase $\phi$, squeezing $r$, transmission $\eta$) is modeled as a zero-mean Gaussian whose variance is the one reported in Eq.(4) of the main text, which we also reproduce here for convenience.

\begin{equation}
\label{eq:sigma2_r_eta}
    \sigma^2(\varphi,r,\eta) = \frac{1}{4}\big(\eta e^{-2r}\cos^2\varphi + (1-\eta) \cos^2\varphi + e^{2r}\sin^2\varphi\big),
    %\sigma^2 = \frac{1}{4}\big( 1 - \eta  +\eta e^{-2r}\cos^2\varphi + e^{2r}\sin^2\varphi\big),
\end{equation}
where $\varphi=\theta-\phi$. \\
In what follows, we describe the 2-parameter adaptive protocol, deriving the model of the corresponding Fisher Information (FI) and, then, its CRB.\\
In particular, in a multiparameter framework where $\vec{y}=\{\phi, r\}$ are the parameters to be estimated, the individual elements of the FI matrix are computed as:
\begin{align}
F_{ij}(\vec{y}| \theta, \eta) = \int_{-\infty}^{\infty} \mathrm{d}x \ p(x|\vec{y}) \left( \frac{\partial}{\partial y_i} \log p(x|\vec{y}, \theta, \eta) \right)
  \left( \frac{\partial}{\partial y_j} \log p(x|\vec{y}, \theta, \eta) \right).
\end{align}

In this case, $\eta$ must be determined by calibration, whereas in the three-parameter estimation protocol it is unknown and therefore included in the data vector $\vec{y}$.\\
If we consider a single-setting experiment in which homodyne data are acquired at a single LO phase, then the resulting FI matrix is singular and the corresponding CRBs are not defined. This is unsurprising: measuring only one quadrature does not provide enough information to distinguish different combinations of $\phi$ and $r$ , i.e., the measurement settings are not tomographically complete. A first achievement of the adaptive protocol we describe is the removal of this ambiguity, which in turn allows one to derive a meaningful multiparameter CRB.\\
As discussed in Sec.~\ref{sec:QFI}, $\phi$ and $r$ are incompatible parameters and cannot be estimated simultaneously with optimal precision. Given that our aim is to choose the optimal setting $\phi_{\mathrm{opt}}$ for estimating $\phi$, we expect $r$ to be estimated with suboptimal precision at that setting. In fact, below we show that the optimal setting for $\phi$ corresponds to the point of minimal sensitivity for $r$. Since the adaptive feedback relies on $\hat{r}$, poor estimation of $r$ also degrades the precision of $\hat{\phi}$; to mitigate this, we design the adaptive protocol accordingly. A natural strategy is to steer the LO phase towards the setting that removes the correlation between the estimation errors, or, equivalently, to drive the off-diagonal element of the FI matrix to zero. This yields
\begin{equation}
   \theta_{\mathrm{opt}} = \hat{\phi} - \phi_{\mathrm{opt}}
= \hat{\phi} -  \arccos\!\left(\frac{e^{2\hat{r}}}{\sqrt{e^{4\hat{r}} + \eta}}\right), 
\end{equation}

where $\hat{\phi}$ and $\hat{r}$ are the current parameter estimate before the adaptive feedback.\\
Keeping in mind this strategy, we now develop a FI matrix model that replicates our multiparameter adaptive protocol. Firstly, we recall the strategy we have implemented experimentally:
\begin{itemize}
    \item \textit{Rough estimation}: the first $M_R =1,200$ homodyne measures are devoted to obtain approximate estimations of the parameters. In particular, we equally divide them in measures with LO phase in $\theta=\{0, \pi/4, \pi/2\}$. This resource allocation in equally spaced settings has the twofold objective of eliminating the aforementioned singularity and provide a sufficiently precise estimation of both parameters.

    \item \textit{Adaptive step}: the estimated parameters $\hat{\phi}$ and $\hat{r}$ are employed to compute $\theta_{\mathrm{opt}}$, and shift the LO phase to this value. In the actual experimental implementation, this step can also be reiterated multiple times to finely adjust the homodyne setting to the optimal value.

    \item \textit{Fine estimation}: as further discussed in Sec.\ref{sec:SMC}, a sufficient amount $M$ of homodyne measures has to be acquired to achieve statistical convergence. In our case, $M=20,000$ is sufficient to ensure such a convergence. Therefore, the remaining $M_F=M- M_R=18,800$ homodyne data are measured within the optimal setting to estimate $\phi$. As we will see, this strategy allows to closely approach the phase QCRB.
\end{itemize}

Therefore, using the definition reported in Eq.\eqref{eq:sigma2_r_eta}, we can model the FI matrix corresponding to the adaptive protocol as a function of the LO phase in the fine estimation step:

\begin{equation}
    \widetilde{\bm{F}}(\phi, r |\theta, \eta) = \underbrace{\left( \frac{1}{3}\bm{F}(\phi, r |0, \eta) + \frac{1}{3}\bm{F}(\phi, r |\pi/4, \eta) + \frac{1}{3}\bm{F}(\phi, r |\pi/2, \eta \right) \frac{M_R}{M}}_{\mathrm{Rough \ estimation}} + \underbrace{\bm{F}(\phi, r | \theta, \eta) \frac{M_F}{M}}_{\mathrm{Fine \ estimation}}.
\end{equation}

The resulting FI matrix is invertible, allowing us to define the corresponding multiparameter CRBs.

\begin{figure*}[htb!]
\centering
\includegraphics[width=0.99\textwidth]{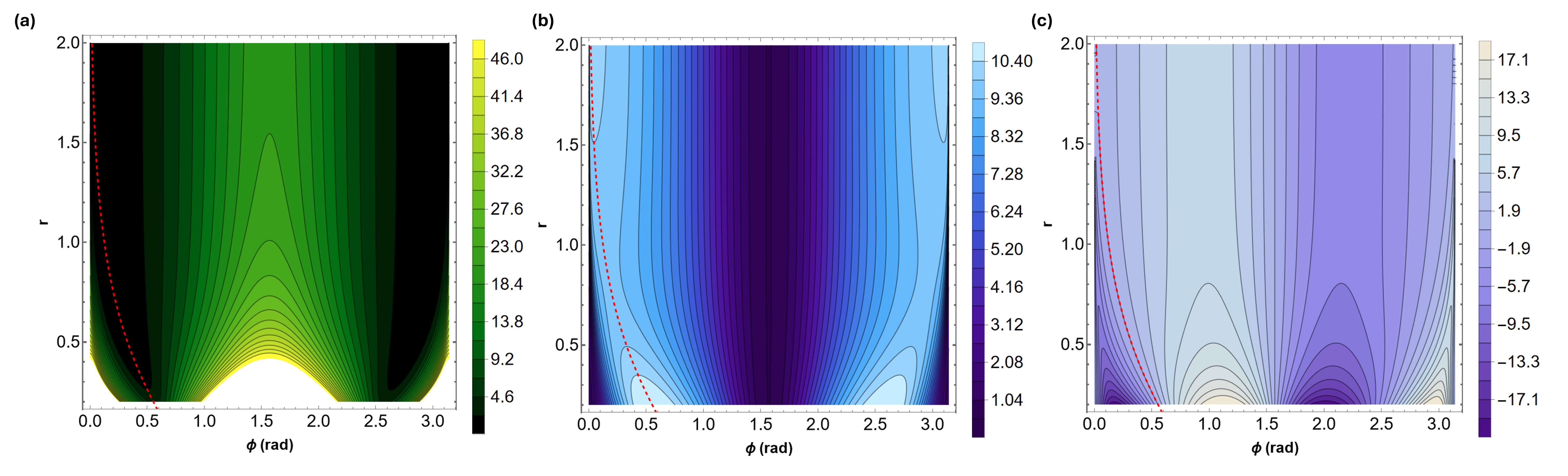}
\caption{\textbf{Inverse of the FI matrix elements in the multiparameter adaptive protocol}. In this figure, we report the behavior of the $\widetilde{\bm{F}}^{-1}$ entries as a function of $\phi \in [0,2 \pi]$ and $r\in [0,2]$. In particular, $[\widetilde{\bm{F}}^{-1}]_{\phi \phi}$, $[\widetilde{\bm{F}}^{-1}]_{rr}$, and $[\widetilde{\bm{F}}^{-1}]_{\phi r}=[\widetilde{\bm{F}}^{-1}]_{r \phi}$ illustrated respectively in panels \textbf{(a)}, \textbf{(b)}, and \textbf{(c)}. In all of them, we report the value of the phase $\phi_{\mathrm{opt}}(r, \eta)$ that eliminates the correlations as a red dashed line.}
\label{fig:FI_r_phi}
\end{figure*}

 In Fig.\ref{fig:FI_r_phi}, we report the behavior of the individual elements of $[\widetilde{\bm{F}}^{-1}]$ as a function of the parameters $\phi$ and $r$. For the sake of clarity, in what follows, we will use the notation $[\widetilde{\bm{F}}^{-1}]_{ij}$ with $i,j \in \{\phi, r\}$ to denote corresponding entries of the inverse FI matrix. In particular, we can notice how the $r$-dependent phase that eliminates the correlations (red dashed lines in Fig.\ref{fig:FI_r_phi}), i.e. $[\widetilde{\bm{F}}^{-1}]_{\phi r}=0$ (right panel) also corresponds to the minimum of the $ [\widetilde{\bm{F}}^{-1}]_{\phi \phi}$ (left panel) and the maximum $[\widetilde{\bm{F}}^{-1}]_{rr}$ (central panel), along the entire range of $r$.

\begin{figure*}[htb!]
\centering
\includegraphics[width=0.95\textwidth]{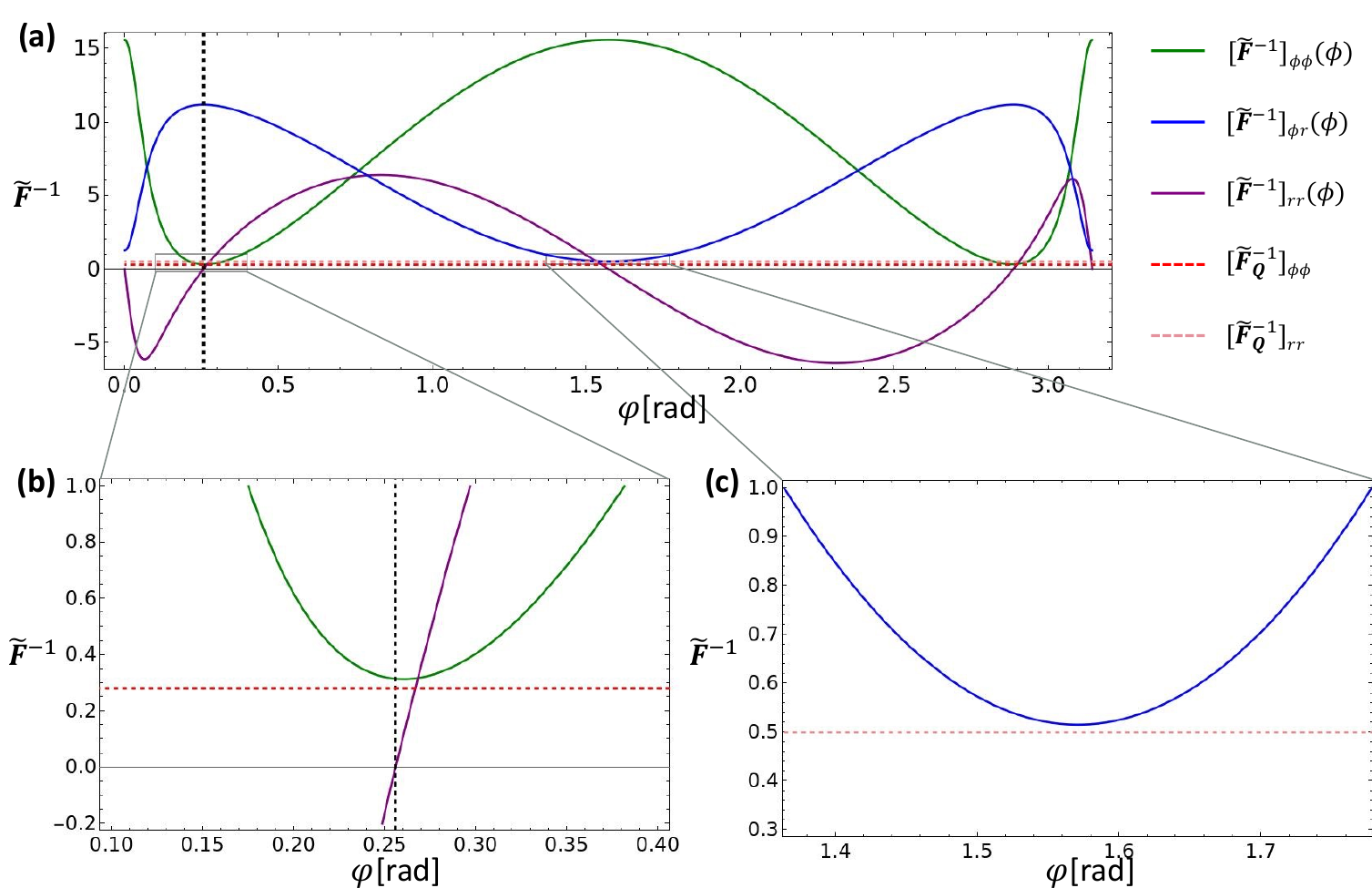}
\caption{\textbf{Adaptive FI matrix model for experimental values of $\bm{r}$ and $\bm{\phi}$.} In this figure, we report the behavior of the adaptive FI matrix entries as a function of $\varphi = \theta - \phi$, when considering the experimentally plausible parameters $\phi=\pi/4$, $r=0.63$, and $\eta=0.85$. In the main panel \textbf{(a)}, we report the $ [\widetilde{\bm{F}}^{-1}]_{\phi \phi}$ (green curve), $ [\widetilde{\bm{F}}^{-1}]_{r r}$ (blue curve), and $[\widetilde{\bm{F}}^{-1}]_{\phi r}$ (purple curve) for the entire range $\varphi \in [0, 2 \pi]$, comparing them to the corresponding QCRBs for the estimation of $\phi$, i.e. $[\bm{F_Q}^{-1}]_{\phi \phi}$ (red dashed line), and for the estimation of $r$, i.e. $[\bm{F_Q}^{-1}]_{r r}$ (orange dashed line). The black dashed line represents the phase $\phi_{\mathrm{opt}}$ that removes the correlation between the estimation errors of $\phi$ and $r$. In the insets \textbf{(b)} and \textbf{(c)}, we report the magnification of the main plot around the optimal settings for estimating $\phi$ and $r$, respectively, again compared to the corresponding QRCBs.
}
\label{fig:FI_matrix_zoom}
\end{figure*}

To have a better understanding of the adaptive FI matrix model we derived, in Fig.\ref{fig:FI_matrix_zoom}, we illustrate the behavior of $[\widetilde{\bm{F}}^{-1}]_{\phi \phi} $, $[\widetilde{\bm{F}}^{-1}]_{r r}$, and $[\widetilde{\bm{F}}^{-1}]_{\phi r}$ when the squeezing level and efficiency are similar to the experimental conditions, i.e. $r=0.63$ and $\eta=0.85$. As previously mentioned, we can see how the setting that maximizes the precision in estimating $\phi$ ($r$), at the same time minimizes the sensibility to $r$ ($\phi$). Additionally, the inset (a) of Fig.\ref{fig:FI_matrix_zoom} shows how the $\phi_{\mathrm{opt}}$ is indeed the one that removes the correlation (purple curve) and optimizes $[\widetilde{\bm{F}}^{-1}]_{\phi \phi}$, which achieves a value quite close to the corresponding QCRB. This consideration is also true for the optimal estimation of $r$, which is detailed in the inset (c) of Fig.\ref{fig:FI_matrix_zoom}.

\begin{figure*}[htb!]
\centering
\includegraphics[width=0.75\textwidth]{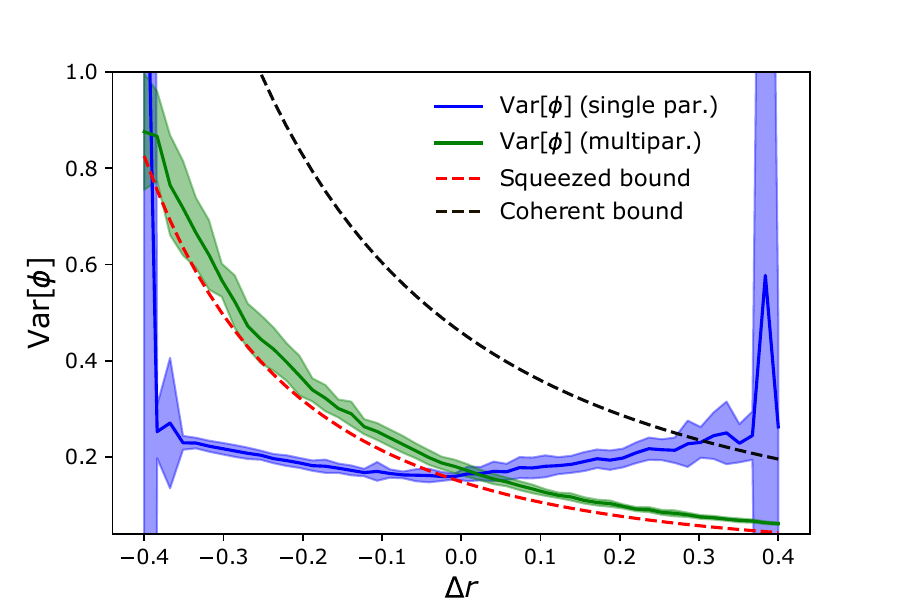}
\caption{\textbf{Behavior of the variance as the squeezing level deviates from the calibration.} In this figure, we report how the phase estimation variance changes when the squeezing level deviates from the original value $r=0.8$ by a quantity $\Delta r \in [-0.4, 0.4]$. In particular, we report the average and the standard deviation (over 50 repetitions) of the simulated estimation variance obtained with the single- (multi-)parameter adaptive protocol as a blue (green) solid curve and shaded area, respectively. We compare them with the corresponding QCRB (red dashed curve) and coherent bound (black dashed curve).
}
\label{fig:r_variabile}
\end{figure*}

Finally, we further motivate the use of the multiparameter approach through numerical simulations
showing that the implemented protocol is inherently robust to variations in the probe squeezing level, whereas the single-parameter approach fails under the same conditions. The results of simulations performed by changing the squeezing by an amount $\Delta r$  are reported in Fig.~\ref{fig:r_variabile}. The plotted results refer to the variance on the phase estimate for a single-parameter estimator that fixes the squeezing at its pre-calibrated value $r$, and our multiparameter adaptive estimator, as the true squeezing is offset by $\Delta r$. 
For the single-parameter approach, even small mismatches brings to unphysical variances or significantly degrade estimation precision. The estimation error approches the bound only when the effective squeezing remains very close to the calibrated value.
On the contrary, the proposed multiparameter adaptive protocol is able to keep the phase estimation variance close to the QCRB for any variation of the squeezing, since it avoids explicit calibration of the probe squeezing and instead infers its value directly from the data. This joint inference makes the protocol robust, suppressing systematic errors from drift and sample variability.

\section{Sequential Monte Carlo}
\label{sec:SMC}

\begin{figure*}[htb!]
\centering
\includegraphics[width=0.6\textwidth]{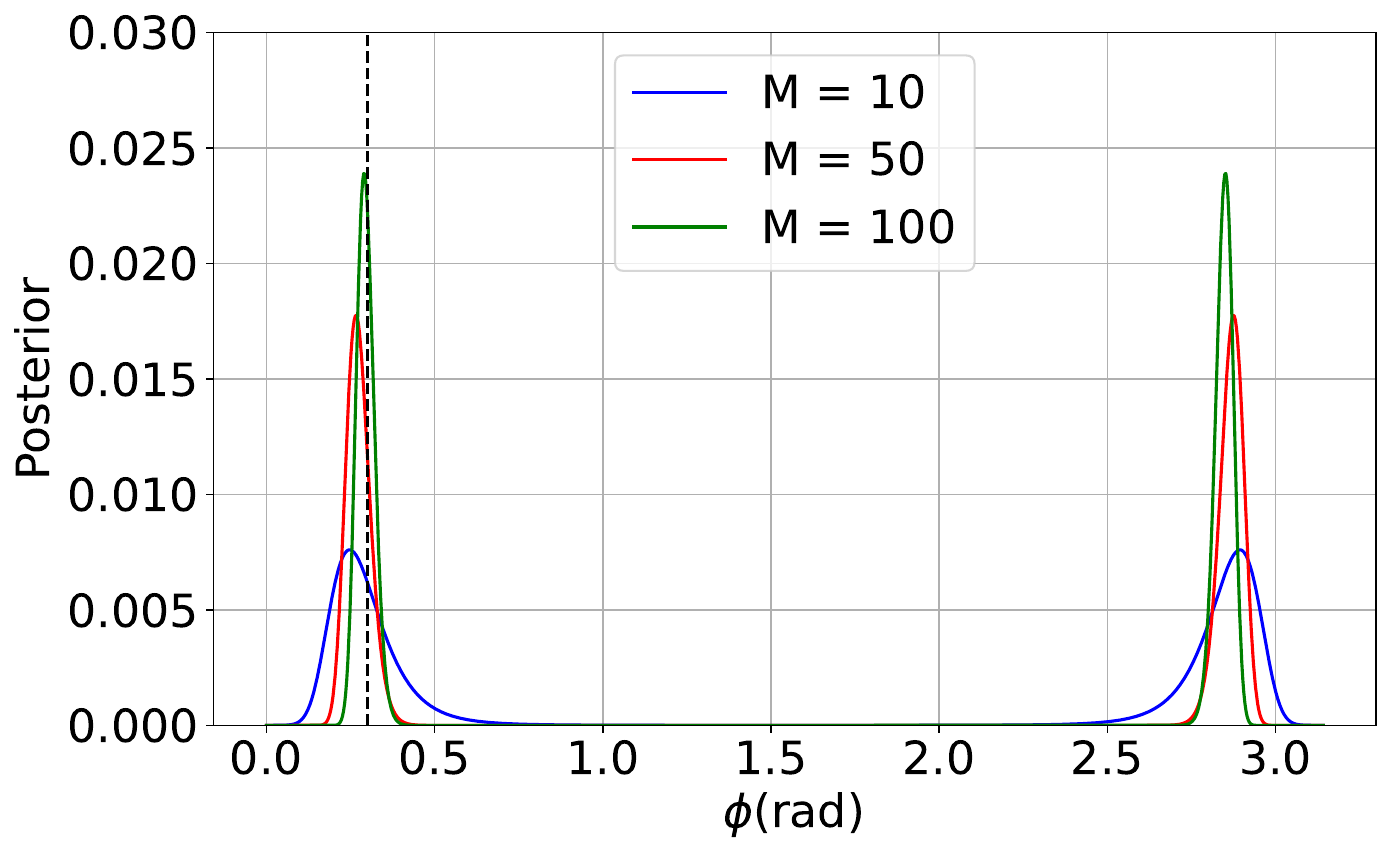}
\caption{\textbf{Simulated Bayesian posterior.} Reconstructed posterior probability distribution of Bayesian phase estimation in range $[0, \pi]$ with $M = 10, 50, 100$ homodyne measures. We choose a trial phase $\phi_{true} = 0.3$ rad and a squeezing parameter $r = 1$.}
\label{fig_posterior}
\end{figure*}

We estimate the unknown phase $\phi$ with a Bayesian procedure that updates a flat prior on $[0,\pi]$ using homodyne data $X=\{x_m\}_{m=1}^M$. Such measurements are employed to reconstruct the single-parameter posterior probability distribution using Bayes' theorem, the posterior probability reconstructed after $M$ measures is reported in Fig.\ref{fig_posterior}.

The reconstructed posterior is updated using the homodyne probability distribution, obtained by marginalizing the Wigner function of the squeezed probe state and setting the phase of the local oscillator (LO) to $\theta$.
Once having reconstructed the posterior probability, we derive the estimate $\hat{\phi}$ for the parameter of interest and its variance. As usual in the Bayesian framework, these are obtained by calculating the mean value and variance of the posterior distribution, respectively:
\begin{equation}
    \hat{\phi} = \int p(\varphi|X)\varphi d\varphi\sim \sum_{k=1}^{n_p} \omega_k(X)\phi_k,
\end{equation}

\begin{equation}
    \text{Var}[\phi] = \int p(\varphi|X)(\hat{\varphi}-\varphi)^2 d\varphi \sim \sum_{k=1}^{n_p} \omega_k(X)(\hat{\phi}-\phi_k)^2.
\end{equation}

In the adopted framework, the computation is accelerated by substituting integrals with discretized sums that have a relevant impact, in particular for multiparameter estimation, where the computation of multidimensional integrals can be costly, affecting the realization of the adaptive protocol.

\begin{figure*}[htb!]
\centering
\includegraphics[width=0.99\textwidth]{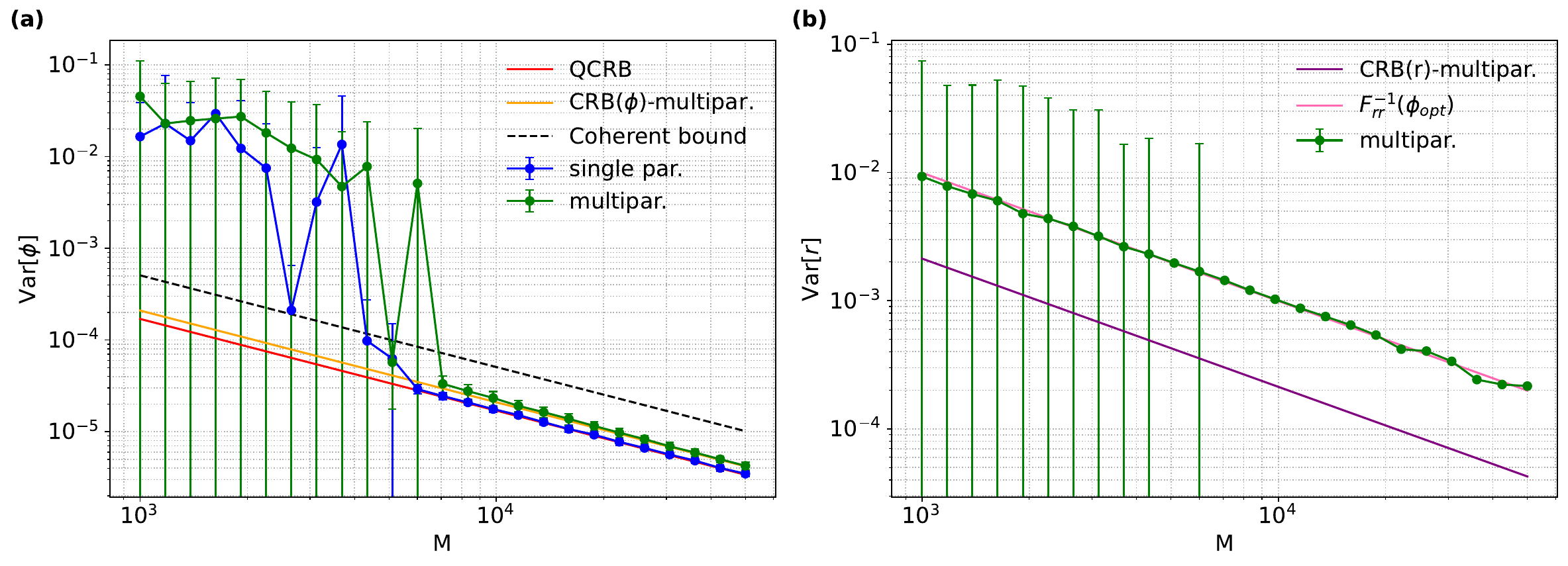}
\caption{\textbf{Scaling of simulated posterior variance with the number of homodyne samples M.} Panel \textbf{(a)} average variance of the phase estimate with the single-parameter Bayesian protocol (blue circles); variance of the phase estimate with the multiparameter Bayesian protocol (green circles). The dashed black line is the ultimate classical bound for coherent states, the red line is the QCRB for squeezed light; the orange line is the CRB for the adopted two-parameter strategy. Panel \textbf{(b)} shows the average variance of the squeezing parameter estimate (green circles). The purple line is the CRB achievable when the protocol is optimized to minimize $\mathrm{Var}(r)$ while the pink line represents the bound obtained with the implemented (phase-optimized) adaptive strategy. Error bars represent the standard deviation over estimates of $20$ different phase values. }
\label{VarM}
\end{figure*}

To verify that the implemented estimator attains the relevant precision limits, we perform simulations of the estimation protocols using the probe parameters $\eta=0.8$ and $r=0.8$. In the single-parameter protocol, we discretize $[0,\pi]$ with $n_p = 10000$ particles and observe the expected $1/M$ variance decay that saturates the QCRB. In the joint $(\phi, r)$ estimate, $n_p = 20000$ total particles were employed, and the trend of the variances on both parameter estimates has been investigated. The simulations have been carried out with the same adaptive protocol described before, whose aim is to minimize the variance on the phase estimate. Studying the performance scaling reported in Fig.\ref{VarM} with the number of homodyne measures is evident how they saturate to the relative elements of the inverse of the FIM computed in the previous section.

